\documentclass[12pt]{article}
\DeclareUnicodeCharacter{2212}{-}
\usepackage{graphicx}
\usepackage{amsmath,amsthm,amsfonts,amscd,amssymb,mathrsfs,cite}
\usepackage[initials]{amsrefs} 
\usepackage{tikz}
\usepackage{adjustbox}
\usepackage{pgfplots}
\usepackage{amscd}   
\usepackage[all]{xy} 
\usepackage{booktabs} 
\usepackage{stmaryrd}
\usepackage{mathtools}
\usepackage{numprint}
\usepackage{epstopdf}
\usepackage{qcircuit}
\usepackage[utf8]{inputenc} 
\usepackage{soul} 
\usepackage{nccmath} 
\usepackage{url}
\usepackage{numprint}
\usepackage{algorithm}
\usepackage{algpseudocode}
\usepackage{subcaption}
\usepackage{rotating}
\usepackage{float}

\usetikzlibrary{patterns}
\tikzset{dottednode/.style={
        dash pattern=on 1.5pt off 4pt, 
        line width=1.9pt
}}
\tikzset{dashednode/.style={dash pattern=on 6pt off 5pt}}

\DeclareMathAlphabet{\mathscr}{OT1}{pzc}{m}{it} 

\usepackage{hyperref}
  \hypersetup{colorlinks=true,citecolor=blue, urlcolor= cyan}

 \AtBeginDocument{%
    \def\MR#1{} 
 }

\usepackage{color}

\newtheorem{definition}{Definition}

\newcommand{\CC}{\mathbb{C}}

\newcommand{\np}[1]{\numprint{#1}}

\theoremstyle{remark}

\usepackage{physics}

\usepackage{authblk}

\definecolor{codegreen}{rgb}{0,0.4,0}
\definecolor{codeorange}{rgb}{0.7,0.5,0}
\definecolor{codegray}{gray}{.7}

\algnewcommand\algorithmicforeach{\textbf{for each}}
\algdef{S}[FOR]{ForEach}[1]{\algorithmicforeach\ #1\ \algorithmicdo}
\algnewcommand{\IIf}[1]{\State\algorithmicif\ #1\ \algorithmicthen}
\algnewcommand{\EndIIf}{\unskip\ \algorithmicend\ \algorithmicif}
\algnewcommand{\IFor}[2]{\State\algorithmicforeach\ #1\ \algorithmicdo #2\ \algorithmicend\ \algorithmicfor}

\title{A new heuristic approach for contextuality degree estimates and its four- to six-qubit portrayals}
\author[$\ast$]{Axel Muller}
\author[$\star$]{Metod Saniga}
\author[$\ast$]{Alain Giorgetti}
\author[$\dagger,\ddagger$]{Frédéric Holweck}
\author[$\dagger$]{Colm Kelleher}

\affil[$\ast$]{Université de Franche-Comté, CNRS, Institut FEMTO-ST, F-25000\nolinebreak[4]  Besançon, France}
\affil[$\star$]{Astronomical Institute of the Slovak Academy of Sciences, SK-05960\nolinebreak[4]  Tatransk\' a Lomnica, Slovakia}
\affil[$\dagger$]{Laboratoire Interdisciplinaire Carnot de Bourgogne, ICB/UTBM, UMR6303, CNRS, Universit\'e de Technologie de Belfort-Montbéliard, F-90010 Belfort Cedex, France}
\affil[$\ddagger $]{Department of Mathematics and Statistics, Auburn University, Auburn\nolinebreak[4] (AL), U.\,S.\,A.}
\date{}
\begin{document}

\maketitle

\abstract{We introduce and describe a new heuristic method for finding an upper bound on the degree of contextuality and the corresponding unsatisfied part of a quantum contextual configuration with three-element contexts (i.e., lines) located in a multi-qubit symplectic polar space of order two. While the previously used method based on a SAT solver was limited to three qubits, this new method is much faster and more versatile, enabling us to also handle four- to six-qubit cases. The four-qubit unsatisfied configurations we found are quite remarkable. That of an elliptic quadric features 315 lines and has in its core three copies of the split Cayley hexagon of order two having a Heawood-graph-underpinned geometry in common. That of a hyperbolic quadric also has 315 lines but, as a point-line incidence structure, is isomorphic to the dual $\mathcal{DW}(5,2)$ of $\mathcal{W}(5,2)$. Finally, an unsatisfied configuration with 1575 lines associated with all the lines/contexts of the four-qubit space contains a distinguished $\mathcal{DW}(5,2)$ centered on a point-plane incidence graph of PG$(3,2)$. The corresponding configurations found in the five-qubit space exhibit a considerably higher degree of complexity, except for a hyperbolic quadric, whose 6975 unsatisfied contexts are compactified around the point-hyperplane incidence graph of PG$(4,2)$. The most remarkable unsatisfied patterns discovered in the six-qubit space are a couple of disjoint split Cayley hexagons (for the full space) and a subgeometry underpinned by the complete bipartite graph $K_{7,7}$ (for a hyperbolic quadric).}

\section{Introduction}\label{sec:intro}

A couple of years ago, de Boutray, Masson and three of the authors~\cite{DHGMS22} introduced the notion of the degree of contextuality of a quantum configuration regarded as a particular sub-geometry of a multi-qubit symplectic polar space of order two and were already able to compute this degree for quadrics of the three-qubit space. In a later work~\cite{muller2023new}, a more efficient SAT-based algorithm was proposed, together with an implementation in C language also able to partially handle the four-qubit case. Moreover, in selected three-qubit configurations, this new algorithm even allowed us to see {\it explicit} point-line geometries formed by the smallest number of unsatisfied contexts, these being -- up to isomorphism -- equal to (i) a set of nine mutually disjoint lines for an elliptic quadric, (ii) a set of 21 lines that can be identified with the edges of the Heawood graph for a hyperbolic quadric and, last but not least, (iii) a ``classical'' copy of the split Cayley hexagon of order two for the configuration comprising all the 315 contexts of the space.

In order to address in a similar vein quantum configurations with more than three qubits, we propose here a new heuristic approach. The approach is much faster and more versatile than the previous one, furnishing either new or considerably improved upper bounds on the degree of contextuality of configurations living in four- to seven-qubit spaces. In addition, in the four- to six-qubit spaces, it also provides a sufficiently detailed geometric understanding of the corresponding unsatisfied parts of contextual configurations under study.

The (rest of the) paper is organized as follows. We start with a brief inventory of the basic concepts and notations in Section~\ref{sec:back}, to be followed, in Section~\ref{sec:heur}, by a detailed description of the new method and a brief tabular listing of the new results. Section~\ref{sec:lobo} offers a chain of combinatorial geometric arguments to ascertain the lower bounds on the degree of contextuality for some specific generic cases. Section~\ref{sec:res} deals with the geometrically-slanted description of the most important new results achieved and provides a fairy detailed illustration of those pertaining to the four-, five- and six- qubit spaces. Finally, in Section~\ref{sec:conc} one recollects the main achievements and outlines possible direction(s) of the future work.

\section{Background concepts and notations}\label{sec:back}

This section revisits the connection between sets of mutually commuting $N$-qubit Pauli operators and the totally isotropic subspaces in the symplectic polar space of rank $N$ and order 2, $\mathcal{W}(2N-1,2)$, as referenced in~\cites{saniga2007multiple,havlicek2009factor,kthas}. The $N$-qubit Pauli group, $\mathcal{P}_N$ is the subset of $GL_{2^N}(\CC)$ composed of the elements $\mathcal{O}$ defined by
\begin{equation}
 \mathcal{O}=sA_1\otimes A_2\dots\otimes A_N, \text{ with } s\in \{\pm 1,\pm i\} \text{ and } A_k \in \{I,X,Y,Z\},
 \label{eq:operator}
\end{equation}
where $X, Y, Z$ are the standard Pauli matrices and $I$ is the corresponding identity matrix. In the sequel, these elements will be shorthanded as $\mathcal{O}=sA_1A_2\cdots A_N$. Given the fact that the Pauli matrices $\{I,X,Y,Z\}$ are expressible through matrix multiplication of $Z$ and $X$ as
\begin{equation}
 \begin{array}{cc}
  I=Z^0.X^0\leftrightarrow (0,0), & X=Z^0.X^1 \leftrightarrow (0,1),\\
  Y=iZ^1.X^1 \leftrightarrow (1,1), & Z=Z^1.X^0 \leftrightarrow (1,0),
 \end{array}
\end{equation}
where the dot `.' stands for the ordinary matrix product, eq.~(\ref{eq:operator}) can be cast into the following form
\begin{equation}
 \mathcal{O}=s(Z^{\mu_1}.X^{\nu_1})(Z^{\mu_2}.X^{\nu_2})\cdots(Z^{\mu_N}.X^{\nu_N}) \text{ with } s\in \{\pm 1,\pm i\}, \mu_i, \nu_j \in \{0,1\}.
\end{equation}
The last expression leads to a surjective map between $\mathcal{P}_N$ and $\mathbb{F}_2^{2N}$ ($\mathbb{F}_2$ being the smallest Galois field):
\begin{equation}
 \pi:\left\{\begin{array}{ccc}
      \mathcal{P}_N & \rightarrow & \mathbb{F}_2^{2N}\\
      \mathcal{O}=s(Z^{\mu_1}.X^{\nu_1})(Z^{\mu_2}.X^{\nu_2})\cdots(Z^{\mu_N}.X^{\nu_N}) & \mapsto& (\mu_1,\mu_2,\dots,\mu_N,\nu_1,\nu_2,\dots,\nu_N).
     \end{array}\right.
     \label{eq:map-vect}
\end{equation}
The center of $\mathcal{P}_N$ is $C(\mathcal{P}_N)=\{\pm I_N,\pm i I_N\}$, where $I_N$ is that particular element defined by  eq.~(\ref{eq:operator}) for which $A_1 = A_2 = \cdots = A_N = I$ and $s=+1$. Therefore, $\mathcal{P}_N/C(\mathcal{P}_N)$ is isomorphic to the additive group $\mathbb{F}_2^{2N}$.
Disregarding the neutral element, we then obtain a correspondence between equivalence classes of $\mathcal{P}_N/C(\mathcal{P}_N)$ and points of PG$(2N-1,2)$, the $(2N-1)$-dimensional projective space over the two-elements field:

\begin{equation}
 \underline{\pi}:\left\{\begin{array}{ccc}
      \mathcal{P}_N/C(\mathcal{P}_N) \backslash C(\mathcal{P}_N) & \rightarrow & {\rm PG}(2N-1,2)\\
      \overline{\mathcal{O}}=\{s(Z^{\mu_1}.X^{\nu_1})(Z^{\mu_2}.X^{\nu_2})\cdots(Z^{\mu_N}.X^{\nu_N})\} & \mapsto& [\mu_1:\mu_2:\dots:\mu_N:\nu_1:\nu_2:\dots:\nu_N].
     \end{array}\right.
     \label{eq:map-proj}
\end{equation}
In $\mathcal{P}_N/C(\mathcal{P}_N)$ two classes $\overline{\mathcal{O}}$ and $\overline{\mathcal{O}'}$ commute if and only if $\sum_{i=1} ^N \mu_i\nu_{i}'+\mu_i'\nu_i=0$, with $\mathcal{O}=s(Z^{\mu_1}.X^{\nu_1})(Z^{\mu_2}.X^{\nu_2})\dots(Z^{\mu_N}.X^{\nu_N})$ and $\mathcal{O}'=s'(Z^{\mu_1'}.X^{\nu_1'})(Z^{\mu_2'}.X^{\nu_2'})\dots(Z^{\mu_N'}.X^{\nu_N'})$ being representatives of either class.
To account for these commutation relations, we introduce on PG$(2N-1,2)$ a non-degenerate symplectic form:
\begin{equation}
 \langle p,q\rangle=\sum_{i=1} ^N p_iq_{N+i}+p_{N+i}q_i,
\end{equation}
with $p=[p_1:p_2:\dots:p_{2N}]$ and $q=[q_1:q_2:\dots:q_{2N}]$. The consequence is that two (classes of pairwise) commuting observables in $\mathcal{P}_N/C(\mathcal{P}_N)$ define a totally isotropic line in PG$(2N-1,2)$ with respect to this symplectic form or, in other words, the commutation relations are now encoded into specific collinearity relations on such a PG$(2N-1,2)$, which also has the proper name:
\begin{definition}
 The space of totally isotropic subspaces\footnote{A totally isotropic subspace of PG$(2N-1,2)$ equipped with a non-degenerate symplectic form is any subspace on which the symplectic form vanishes identically; a totally isotropic subspace of maximal (projective) dimension $N-1$ is called a generator of $\mathcal{W}(2N-1,2)$.} of {\rm PG}$(2N-1,2)$ equipped with a non-degenerate symplectic form $\langle,\rangle$ is called the \emph{symplectic polar space of rank $N$ and order two}, usually denoted as $\mathcal{W}(2N-1,2)$.
\end{definition}
\noindent
The smallest non-trivial space, $N=2$, often called the \emph{doily}, is remarkable in the sense that it is also the smallest thick generalized quadrangle~\cite{p-t09} and the sole out of $\np{245342}$ $15_3$-configurations that is triangle-free~\cite{rich}.

Viewed as a point-line incidence structure, $\mathcal{W}(2N-1,2)$ contains
\begin{equation}
|\mathcal{W}|_p =  4^N - 1  
\label{eq:w-p}
\end{equation}
points and
\begin{equation}
|\mathcal{W}|_l  = (4^N - 1)(4^{N-1} - 1)/3
\label{eq:w-l}
\end{equation}
lines, with three points per line and $4^{N-1} - 1$ lines through a point. Given a point of $\mathcal{W}(2N-1,2)$, then the lines, planes, \ldots, and generators passing through this point form a geometry isomorphic to $\mathcal{W}(2N-3,2)$. Moreover, $\mathcal{W}(2N-1,2)$ contains two specific types of subgeometries, hyperbolic and elliptic quadrics, that are of particular importance regarding the study of contextual configurations~\cite{DHGMS22}:
\begin{enumerate}
\item {\bf Hyperbolic Quadric} $\mathcal{Q}^{+}(2N-1,2)$: For $N \geq 1$, it is defined by the standard canonical equation:

   \[
   x_1x_{N+1}+x_2x_{N+2}+\cdots+x_{N}x_{2N}=0.
   \]

   Each $\mathcal{Q}^{+}(2N-1,2)$ is endowed with $(2^{N-1} + 1)(2^{N} -1)$ points and there are $(2^{N-1} + 1)(2^{N} -1) + 1$ copies of them in $\mathcal{W}(2N-1,2)$.

\item {\bf Elliptic Quadric} $\mathcal{Q}^{-}(2N-1,2)$: For $N \geq 2$, it consists of all points and subspaces in $\mathcal{W}(2N-1,2)$ that satisfy the standard equation:

   \[
   f(x_1,x_{N+1})+x_2x_{N+2}+\cdots+x_{N}x_{2N} = 0,
   \]

   where $f$ is an irreducible polynomial over $\mathbb{F}_2$. Each $\mathcal{Q}^{-}(2N-1,2)$ contains $(2^{N-1} - 1)(2^{N} + 1)$ points and in $\mathcal{W}(2N-1,2)$ there are $(2^{N-1} - 1)(2^{N} + 1) + 1$ copies of these configurations.
\end{enumerate}
In what follows, we will select from each equivalence class of $\mathcal{P}_N/C(\mathcal{P}_N)$ a single representative, namely the canonical one ($s=1$), to label a particular point of $\mathcal{W}(2N-1,2)$.
A canonical observable $\mathcal{O}$ is either symmetric ($\mathcal{O}^{{\rm T}} = \mathcal{O}$), or skew-symmetric ($\mathcal{O}^{{\rm T}} = - \mathcal{O}$).
It then can be shown that given a canonical observable $\mathcal{O}$,  the set of symmetric canonical observables commuting with $\mathcal{O}$ together with the set of skew-symmetric observables not
commuting with $\mathcal{O}$ lie on a quadric of  $\mathcal{W}(2N-1,2)$, this quadric being hyperbolic (resp. elliptic) if $\mathcal{O}$ is symmetric (resp. skew-symmetric). We call this associated observable the {\it index} of a quadric and can express it, if appropriate, in a subscript, $\mathcal{Q}^{\pm}_{\mathcal{O}}(2N-1,2)$, noting that there exists a particular hyperbolic quadric associated with $I_N$, $\mathcal{Q}^{+}_{I_N}(2N-1,2).$ Also, when referring in the sequel to $\mathcal{W}(2N-1,2)$ we will always have in mind the $\mathcal{W}(2N-1,2)$ with its
points being labelled by canonical $N$-qubit observables as dictated by eqs.~(\ref{eq:map-vect}) and~(\ref{eq:map-proj}). 
Moreover, a line (or any other linear subspace) of such a multi-qubit $\mathcal{W}(2N-1,2)$ will be called \emph{positive} or \emph{negative} according as the (ordinary) product of
the observables located in it is $+ I_N$ or $- I_N$, respectively.

Next, given an even-dimensional projective space over $\mathbb{F}_2$, PG$(2N,2)$, $N > 1$, a parabolic quadric in this space, $\mathcal{Q}(2N,2)$, is defined by the following canonical equation
\[
   x_1x_{N+1}+x_2x_{N+2}+\cdots+x_{N}x_{2N} + x_{2N+1}^{2} =0.
   \]
Any such quadric has a notable property that all its tangent hyperplanes pass through the same point (see, e.\,g.,~\cites{hir-thas,k-s23,bdb16}), which is usually referred to as the \emph{nucleus} of the quadric.   
Another remarkable well-known fact is (see, e.\,g.,~\cites{hir,hir-thas}) that 
\[
\mathcal{W}(2N-1,2) \cong \mathcal{Q}(2N,2).
\]
In order to distinguish between these two cases, we shall call -- following the notation of~\cite{sbhg21} -- the
$\mathcal{W}(2N-1,2)$ embedded into PG$(2N-1,2)$ a linear $\mathcal{W}(2N-1,2)$,
whereas the space represented by $\mathcal{Q}(2N,2)$ in PG$(2N,2)$ will be referred to as
a quadratic $\mathcal{W}(2N-1,2)$. It is worth mentioning that the intersection of a 
$\mathcal{Q}^{+}(2N-1,2)$ and a $\mathcal{Q}^{-}(2N-1,2)$ is isomorphic to a
quadratic $\mathcal{W}(2N - 3,2)$ (see, e.\,g., \cite{VL10}).

We shall also employ the notion of a finite \emph{point-line incidence structure} ${C} = ({P},{L}, \in \nolinebreak)$, where ${P}$ and ${L}$ are, respectively, finite sets of points and lines and where $\in$ is a binary relation between ${P}$ and ${L}$, indicating which point-line pairs are incident; the number of lines of ${C}$ incident with a point of ${C}$ will be called the \emph{degree} of the point. Any distinguished subset of ${P}$ such that a line of ${L}$ is either fully contained in it or shares with it just a single point is called a \emph{geometric hyperplane} of ${C}$~\cite{ronan}.

Given $\mathcal{W}(5,2)$, its dual space, $\mathcal{DW}(5,2)$, as a point-line incidence structure, has for its points the 135 planes of $\mathcal{W}(5,2)$ and for its lines the 315 lines of $\mathcal{W}(5,2)$, the incidence being containment (see, e.\,g., \cite{pralle}). As a plane of $\mathcal{W}(5,2)$ has seven lines, there are seven lines passing through a point of $\mathcal{DW}(5,2)$; and as there are three planes sharing a line in $\mathcal{W}(5,2)$, each line of $\mathcal{DW}(5,2)$ features three points. Hence, $\mathcal{DW}(5,2)$ is a specific $(135_7, 315_3)$-configuration. We also mention that $\mathcal{DW}(5,2)$ contains 63 $\mathcal{W}(3,2)$'s, three sharing a line and seven through a point; each such doily is, as a point-line incidence structure,  represented by 15 planes and 15 lines passing via the same point of $\mathcal{W}(5,2)$, the incidence being containment.

Another relevant geometry associated with  $\mathcal{W}(5,2)$ is the split Cayley hexagon of order two, $\mathcal{H}$, which is a $(63_3)$-configuration whose smallest polygons are hexagons (see, e.\,g., \cite{psm}). It is embeddable into $\mathcal{W}(5,2)$ in two different ways, called classical and skew~\cite{cool}. A classically-embedded hexagon, $\mathcal{H_C}$, possesses in $\mathcal{W}(5,2)$ a much greater symmetry than its skew-embedded cousin, $\mathcal{H_S}$. To see this, let us call a point of $\mathcal{H}$ embedded in $\mathcal{W}(5,2)$ \emph{planar} if all the three lines passing through it lie in the same plane of $\mathcal{W}(5,2)$. An $\mathcal{H_C}$ is characterized by the fact that each of its points is planar. In an $\mathcal{H_S}$, only 15 points have this property; they are situated on three pairs of concurrent lines, the three points of concurrence lying themselves on a line -- the latter called the \emph{axis} of $\mathcal{H_S}$. There are altogether 120 $\mathcal{H_C}$'s and 7560 $\mathcal{H_S}$'s in a $\mathcal{W}(5,2)$, as first ascertained by a computer-aided search in~\cite{holweck2022three} and later given a computer-free, geometric-combinatorial substantiation in~\cite{Sanigaetal23}.

We will also encounter several distinguished graphs. Here we define two (families) of them, which are both bipartite and closely related to each other. The first is a point-hyperplane
incidence graph of PG$(d,2)$, $d \geq 1$. The vertices of this graph are both points and hyperplanes (i.\,e., subspaces of maximal dimension) of PG$(d,2)$, where a vertex represented by a point is connected to a vertex represented by a hyperplane iff the point belongs to the hyperplane. The other is the Haar graph of a positive integer 
$n$, $H(n)$ (see, e.\,g, \cite{wss}), defined as follows. Let us consider a binary representation of $n$,
\[
   n = b_{k-1} 2^{k-1} + b_{k-2} 2^{k-2} + \cdots + b_1 2^1 +b_0,
   \]
where $(b_{k-1}, b_{k-2},\ldots,b_1,b_0)$, with $b_{k-1}=1$, is the binary vector of $n$.
A graph $H(n)$ has two disjoint vertex  sets $u_i$ and $v_i$, $i=0, 1, 2, \ldots, k-1$, with $u_i$ being adjacent to $v_{i+j}$ if and only if $b_j=1$ (mod $k$). Interestingly, each point-hyperplane
incidence graph of PG$(d,2)$ is also a Haar graph; for example, the famous Heawood graph, {\it aka} the point-line incidence graph of PG$(2,2)$, is isomorphic to
$H(69)$.

Finally, in the context of this paper, a \emph{quantum configuration} is a pair $(O,C)$ where $O=\{M_1,\ldots,M_p\}$ is a set of  $p = |O|$ canonical Pauli observables, here identified with points of a multi-qubit $\mathcal{W}(2N-1,2)$, and $C=\{c_1,\ldots,c_l\}$ is a set of $l = |C|$ \emph{contexts}, here limited to lines of $\mathcal{W}(2N-1,2)$. Its \emph{incidence matrix} $A \in \mathbb{F}_2^{l \times p}$ is defined by $A_{i,j} = 1$ if the $i$-th context $c_i$ contains the $j$-th observable $M_j$. Otherwise, $A_{i,j} = 0$. Its \emph{valuation vector} $E \in \mathbb{F}_2^{l}$ is defined by $E_i = 0$ if the line $c_i$ is positive and $E_i = 1$ if it is negative. Then the \emph{degree of contextuality} of $(O,C)$ is $d$ defined~\cite{DHGMS22} by
\begin{equation}
d=d_H(E,\text{Im}(A)),
\label{Hamming}
\end{equation} where $d_H$ is the Hamming distance on the vector space $\mathbb{F}_2^l$.

Given a quantum configuration ${K} = (O,C)$, one can associate with it a configuration $\widetilde{{K}}$ that is geometrically identical with ${K}$, but has its observables replaced by $+1$'s and $-1$'s by an \emph{assignment function} $a : O \rightarrow \{-1,+1\}$ and the sign of each context $c \in C$ replaced by the product $\Pi_{M \in c}~a(M)$ of these $+1$'s and $-1$ over its members.
Given the ${K}$ and a $\widetilde{{K}}$, the configuration with the same points and consisting of those contexts of ${K}$ that have different signs than the corresponding lines in $\widetilde{{K}}$ is called an \emph{unsatisfied configuration} of ${K}$ and is denoted as $\widetilde{{K}}^{{\tt uns}}$.

In what follows, we will use special symbols ${E}^{{\tt uns}}_{N}$, ${H}^{{\tt uns}}_{N}$ and ${F}^{{\tt uns}}_{N}$, with $N > 1$, for an unsatisfied configuration of $\mathcal{Q}^{-}(2N-1,2)$, $\mathcal{Q}^{+}(2N-1,2)$ and of the configuration consisting of all lines of $\mathcal{W}(2N-1,2)$, respectively. For a given ${K}$ there are, of course, a (large) number of $\widetilde{{K}}$'s differing from each other in the distribution of $+1$'s and $-1$'s across their points and so, in general, several associated (mutually non-isomorphic) $\widetilde{{K}}^{{\tt uns}}$'s.

From Formula~(\ref{Hamming}) it follows that the degree de contextuality $d_{{K}}$ of the quantum configuration ${K}$ is also the minimal number of  unsatisfiable contexts in it, in other words the number of contexts in any unsatisfied configuration $\widetilde{{K}}^{{\tt uns}}$ of ${K}$ of minimal size. So, to determine the contextuality degree $d_{{K}}$ of ${K}$, it is sufficient to
find out such an unsatisfied configuration $\widetilde{{K}}^{{\tt uns}}$ that has the {\it smallest} possible number of contexts, with $d_{{K}}$ then being this number of contexts. Obviously, the particular $\widetilde{{K}}$ all of whose
points are labelled by $+1$'s gives $\widetilde{{K}}^{{\tt uns}}$ whose contexts correspond solely to the negative contexts of ${K}$, thus providing the number of negative contexts of ${K}$ as a natural {\it upper} bound for  $d_{{K}}$. Hence, our primary, and almost exclusively computer-aided, effort will be focused on discerning those $\widetilde{{K}}^{{\tt uns}}$'s that have a (possibly much) smaller number of contexts than the number of negative contexts in  ${K}$ and which thus furnish (considerably) reduced upper bounds on the value of $d_{{K}}$.

\section{Novel computer-aided heuristics for ascertaining an upper bound on the contextuality degree}\label{sec:heur}

\subsection{Contextuality of quantum configurations}

A quantum configuration ${K}$ is contextual if its degree of contextuality $d$, as defined by formula~(\ref{Hamming}), is different from zero. The problem of finding $d$ can be formulated as an optimization problem, namely the maximization of satisfied contexts in ${K}$. As already described in Section~\ref{sec:back}, it can be addressed by looking for an associated unsatisfied configuration $\widetilde{{K}}$ that has the maximal possible number of lines with the same sign as in ${K}$. The conjunction of these sign constraints can be expressed as a system of linear equations over $\mathbb{F}_2$~\cite{muller2023new}. The problem of maximizing the number of valid equations of this kind is known as Max-E3-Lin-2~\cite{Hastad01}, when each equation contains exactly three variables. When the sign constraints are encoded as XOR clauses, the optimization problem is the MAX-XOR-SAT problem~\cite{Min02}. Either way, in the general case, solving these problems is known to be in the class APX of problems for which there is a polynomial-time algorithm able to find an approximation of the solution (within a given performance ratio)~\cite{Min02}.

Without leveraging properties specific to the structure of the constraints coming from a quantum configuration, finding a solution to these problem is as complex as SAT solving. SAT solvers have enabled us to determine the contextuality degree for quantum geometries up to three qubits~\cite{muller2023new}. However, due to the exponential growth of the computing time of the solver with respect to the size of these geometries, this approach is unfeasible for higher-rank symplectic polar spaces.

Furthermore, it has recently been shown~\cites{TLC22,mg24} that given an abstract configuration endowed with specific observables and contexts (for example, the doily endowed with two-qubit observables), any other admissible quantum assignment of this configuration (e.\,g., any doily located in the multi-qubit $\mathcal{W}(2N-1,2)$, for any $N>2$) will yield {\it  the same} degree of contextuality. Hence, when checking for contextuality of a particular geometry it suffices to consider its simplest (i.e., with lowest number of qubits) quantum assignment.

To find approximations of the contextuality degree for each geometry, we have first experimented off-the-shelf heuristic-based methods. Google OR-tools~\cite{ortools} find optimal solutions for up to three qubits, by using local and core-based (which focus on solving a critical subset of the problem to enhance computational efficiency) approaches, and provide some better bounds for four qubits than a SAT solver alone within a reasonable time frame. However they fail to provide any useful results for five qubits and more. 

\subsection{Optimization algorithm}

For geometries with more qubits, we suspected that specific heuristics based on some particular properties of the geometry could yield improved bounds. Since these geometries often exhibit symmetries, we designed and present in this section a heuristic grouping variables by the number of unsatisfied constraints they are part of.

\begin{algorithm}[tbh!]
\caption{Optimize Hamming Distance\label{ohdAlgo}}
\begin{algorithmic}[1]
\Function{optimize\_hamming\_distance}{$O,C,E$}
    \For{$o \in O$}
      \State $a[o] \gets 1$ \label{lineInitA}
      \State $\textit{uns}[o] \gets$ number of contexts containing $o$ unsatisfied by $a$ \label{lineInitUns}
    \EndFor
    \State $\textit{min\_a} \gets a$
    \For{$i \gets 1$ to $\textit{MAX\_ITERATIONS}$}
        \For{$o \in O$}
            \If{$\textit{uns}[o] > \theta \times \text{max(\textit{uns})} \text{ and } \text{random()} > \gamma$} \label{lineChoice}
                \State $a[o] \gets - a[o]$ \label{lineSwap}
                \For{\textbf{each} context $c$ such that $o \in c$} \label{lineUnsUpdateBegin}
                    \State $\textit{sign} \gets \Pi_{m \in c}~a[m]$ \label{lineSign}
                    \For{\textbf{each} $m \in c$} 
                        \State $\textit{uns}[m] \gets \textit{uns}[m] - E_c \times \textit{sign}$ \label{lineUnsUpdate}
                    \EndFor
                \EndFor \label{lineUnsUpdateEnd}

            \EndIf
        \EndFor
        \IIf{$d_H(a,E) < d_H(\textit{min\_a},E)$}{~$\textit{min\_a} \gets a$}
    \EndFor
    \State \textbf{return} $\textit{min\_a}$
\EndFunction
\end{algorithmic}
\end{algorithm}

Given a quantum configuration ${K} = (O,C)$ and its valuation vector $E$, Algorithm~\ref{ohdAlgo} performs a stochastic local search for its optimum Hamming distance, as expressed by Formula~(\ref{Hamming}). Its local variables $a$ and \textit{min\_a} of type $O \rightarrow \{-1,+1\}$ are two assignment functions. The second array \textit{min\_a} stores the best assignment currently found by the algorithm. The first array $a$ stores a candidate for a better solution. It is initialized to 1 for each $o \in O$, on Line~\ref{lineInitA}. For each $o \in O$ the local variable $\textit{uns} : O \rightarrow \mathbb{N}$ stores the number $\textit{uns}[o]$ of contexts containing $o$ that are left unsatisfied by $a$. Therefore, it is initialized by the number of negative contexts through $o$, on Line~\ref{lineInitUns}.

To illustrate the algorithm in a graphical way, let us consider -- as a particularly apt example -- a two-spread of a quadratic four-qubit doily, as depicted in Figure~\ref{fig:algo-table}. Let us recall (see, for example, \cite{polster}) that a two-spread of a doily is a $(15_2,10_3)$-configuration that we get if we remove from the doily any set of five pairwise disjoint lines. It was already proved~\cite{muller2023new} that a multi-qubit two-spread is a contextual quantum configuration whose degree of contextuality is one.  Our selected four-qubit two-spread features five negative lines, namely those highlighted in red. The three two-spread assignments (a), (b) and (c) and the corresponding rows in the table incorporated in Figure~\ref{fig:algo-table} show the three steps of an execution of the algorithm.

The algorithm is parameterized by the \emph{threshold} $\theta$ on the number of unsatisfied contexts an observable is in for it to have its value possibly changed, relatively to the maximal number \text{max(\textit{uns})} of unsatisfied contexts in the current assignment. By swapping the sign (Line~\ref{lineSwap}) of most of the observables which satisfy this criterion (Line~\ref{lineChoice}), the algorithm can significantly reduce the Hamming distance of the whole configuration.
 For instance, in the example of execution depicted in Figure~\ref{fig:algo-table}, the value $0.8$ has been chosen for $\theta$. In the initial state (a), it is visible in the first row of the table and in the first subfigure that the maximal number of unsatisfied contexts per observable is two, which means that only observables pertaining to two unsatisfied contexts will be considered since $2 > 0.8\times 2$, but $1 < 0.8\times 2$ and $0 < 0.8\times 2$.

The second parameter of the algorithm is a \emph{sign flip probability} $\gamma$ for selected observables to have their signs flipped. On Line~\ref{lineChoice} this value is compared to the result of the $\text{random}()$ function that uniformly returns a random real number between 0 and 1. This addition of randomness significantly reduces the observed issue of cycling over a too small subset of visited assignments. In the example in Figure~\ref{fig:algo-table} the chosen value for $\gamma$ is $0.7$.

Every time the sign $a[o]$ of an observable $o$ is flipped (Line~\ref{lineSwap}), the number $\textit{uns}[m]$ of unsatisfied contexts of each observable $m$ sharing a context $c$ with $o$ is updated (Lines~\ref{lineUnsUpdateBegin}-\ref{lineUnsUpdateEnd}). First, the new sign of the context $c$ is computed on Line~\ref{lineSign}. Then, the number of unsatisfied contexts $\textit{uns}[m]$ is decremented if this sign is the expected one $E_c$, meaning that the context is now satisfied, and incremented otherwise (Line~\ref{lineUnsUpdate}).

\begin{figure}[H]
\centering
\begin{subfigure}{0.45\textwidth}
\begin{tikzpicture}[every plot/.style={smooth, tension=2},
    scale=3.2,
    every node/.style={scale=0.69,circle,draw=black,fill=white,minimum size=1.0cm}
    ]
  \coordinate (a) at (0,1.0);
  \coordinate (b) at (0,-0.80);
  \coordinate (c) at (0,-0.40);
  \coordinate (d) at (-0.95,0.30);
  \coordinate (e) at (0.76,-0.24);
  \coordinate (f) at (0.38,-0.12);
  \coordinate (g) at (-0.58,-0.80);
  \coordinate (h) at (0.47,0.65);
  \coordinate (i) at (0.23,0.32);
  \coordinate (j) at (0.58,-0.80);
  \coordinate (k) at (-0.47,0.65);
  \coordinate (l) at (-0.23,0.32);
  \coordinate (m) at (0.95,0.30);
  \coordinate (n) at (-0.76,-0.24);
  \coordinate (o) at (-0.38,-0.12);
\draw (d) -- (n) -- (g);
\draw [thick,dashed,red,style=double] (a) -- (k) -- (d);
\draw (g) -- (b) -- (j);
\draw (j) -- (e) -- (m);
\draw (m) -- (h) -- (a);

\draw plot coordinates{(i) (e) (c)};
\draw [thick,dashed,red,style=double] plot coordinates{(c) (n) (l)};
\draw [thick,dashed,red,style=double] plot coordinates{(f) (b) (o)};
\draw [thick,dashed,red,style=double] plot coordinates{(l) (h) (f)};
\draw [thick,dashed,red,style=double] plot coordinates{(o) (k) (i)};

\node[align=center] at (a) {YXYI\\\textcolor{codegreen}{+1}};
\node[align=center] at (b) {XZYY\\\textcolor{codegreen}{+1}};
\node[align=center] at (c) {ZYIY\\\textcolor{codegreen}{+1}};
\node[align=center] at (d) {ZXZY\\\textcolor{codegreen}{+1}};
\node[align=center] at (e) {IXXZ\\\textcolor{codegreen}{+1}};
\node[align=center] at (f) {ZIYX\\\textcolor{codegreen}{+1}};
\node[align=center] at (g) {ZIYZ\\\textcolor{codegreen}{+1}};
\node[align=center] at (h) {IZZY\\\textcolor{codegreen}{+1}};
\node[align=center] at (i) {ZZXX\\\textcolor{codegreen}{+1}};
\node[align=center] at (j) {YZIX\\\textcolor{codegreen}{+1}};
\node[align=center] at (k) {XIXY\\\textcolor{codegreen}{+1}};
\node[align=center] at (l) {ZZXZ\\\textcolor{codegreen}{+1}};
\node[align=center] at (m) {YYXY\\\textcolor{codegreen}{+1}};
\node[align=center] at (n) {IXXX\\\textcolor{codegreen}{+1}};
\node[align=center] at (o) {YZIZ\\\textcolor{codegreen}{+1}};

\node[align=center] at (f) {ZIYX\\\Large\textcircled{\normalsize\textcolor{codegreen}{+1}}};
\node[align=center] at (k) {XIXY\\\Large\textcircled{\normalsize\textcolor{codegreen}{+1}}};
\node[align=center] at (o) {YZIZ\\\Large\textcircled{\normalsize\textcolor{codegreen}{+1}}};
\end{tikzpicture}
\caption{}
\end{subfigure}
\begin{subfigure}{0.4\textwidth}

\begin{tikzpicture}[every plot/.style={smooth, tension=2},
    scale=3.2,
    every node/.style={scale=0.69,circle,draw=black,fill=white,minimum size=1.0cm}
    ]
\draw (d) -- (n) -- (g);
\draw [red,style=double] (a) -- (k) -- (d);
\draw (g) -- (b) -- (j);
\draw (j) -- (e) -- (m);
\draw (m) -- (h) -- (a);

\draw plot coordinates{(i) (e) (c)};
\draw [thick,dashed,red,style=double] plot coordinates{(c) (n) (l)};
\draw [thick,dashed,red,style=double] plot coordinates{(f) (b) (o)};
\draw [red,style=double] plot coordinates{(l) (h) (f)};
\draw [thick,dashed,red,style=double] plot coordinates{(o) (k) (i)};

\node[align=center] at (a) {YXYI\\\textcolor{codegreen}{+1}};
\node[align=center] at (b) {XZYY\\\textcolor{codegreen}{+1}};
\node[align=center] at (c) {ZYIY\\\textcolor{codegreen}{+1}};
\node[align=center] at (d) {ZXZY\\\textcolor{codegreen}{+1}};
\node[align=center] at (e) {IXXZ\\\textcolor{codegreen}{+1}};
\node[align=center] at (f) {ZIYX\\\textcolor{red}{-1}};
\node[align=center] at (g) {ZIYZ\\\textcolor{codegreen}{+1}};
\node[align=center] at (h) {IZZY\\\textcolor{codegreen}{+1}};
\node[align=center] at (i) {ZZXX\\\textcolor{codegreen}{+1}};
\node[align=center] at (j) {YZIX\\\textcolor{codegreen}{+1}};
\node[align=center] at (k) {XIXY\\\textcolor{red}{-1}};
\node[align=center] at (l) {ZZXZ\\\textcolor{codegreen}{+1}};
\node[align=center] at (m) {YYXY\\\textcolor{codegreen}{+1}};
\node[align=center] at (n) {IXXX\\\textcolor{codegreen}{+1}};
\node[align=center] at (o) {YZIZ\\\textcolor{red}{-1}};
\node[align=center] at (o) {YZIZ\\\Large\textcircled{\normalsize\textcolor{red}{-1}}};
\end{tikzpicture}
\caption{\hspace*{-0.8cm}}
\end{subfigure}

\hspace*{-2cm}
\begin{subfigure}{0.3\textwidth}
\begin{tikzpicture}[every plot/.style={smooth, tension=2},
    scale=3.2,
    every node/.style={scale=0.69,circle,draw=black,fill=white,minimum size=1.0cm}
    ]
\draw (d) -- (n) -- (g);
\draw [red,style=double] (a) -- (k) -- (d);
\draw (g) -- (b) -- (j);
\draw (j) -- (e) -- (m);
\draw (m) -- (h) -- (a);

\draw plot coordinates{(i) (e) (c)};
\draw [thick,dashed,red,style=double] plot coordinates{(c) (n) (l)};
\draw [red,style=double] plot coordinates{(f) (b) (o)};
\draw [red,style=double] plot coordinates{(l) (h) (f)};
\draw [red,style=double] plot coordinates{(o) (k) (i)};

\node[align=center] at (a) {YXYI\\\textcolor{codegreen}{+1}};
\node[align=center] at (b) {XZYY\\\textcolor{codegreen}{+1}};
\node[align=center] at (c) {ZYIY\\\textcolor{codegreen}{+1}};
\node[align=center] at (d) {ZXZY\\\textcolor{codegreen}{+1}};
\node[align=center] at (e) {IXXZ\\\textcolor{codegreen}{+1}};
\node[align=center] at (f) {ZIYX\\\textcolor{red}{-1}};
\node[align=center] at (g) {ZIYZ\\\textcolor{codegreen}{+1}};
\node[align=center] at (h) {IZZY\\\textcolor{codegreen}{+1}};
\node[align=center] at (i) {ZZXX\\\textcolor{codegreen}{+1}};
\node[align=center] at (j) {YZIX\\\textcolor{codegreen}{+1}};
\node[align=center] at (k) {XIXY\\\textcolor{red}{-1}};
\node[align=center] at (l) {ZZXZ\\\textcolor{codegreen}{+1}};
\node[align=center] at (m) {YYXY\\\textcolor{codegreen}{+1}};
\node[align=center] at (n) {IXXX\\\textcolor{codegreen}{+1}};
\node[align=center] at (o) {YZIZ\\\textcolor{codegreen}{+1}};
\end{tikzpicture}
\caption{\hspace*{-2.5cm}}
\end{subfigure}
\vspace{1cm}

\setlength{\tabcolsep}{8pt}
\begin{tabular}{c|c|c|c|c|c|c|c|c|c|c|c|c|c|c|c|| c}
    &
    \rotatebox{-90}{YXYI} & \rotatebox{-90}{XZYY}& \rotatebox{-90}{ZYIY} & \rotatebox{-90}{ZXZY} & \rotatebox{-90}{IXXZ} & \rotatebox{-90}{ZIYX} & \rotatebox{-90}{ZIYZ} & \rotatebox{-90}{IZZY} & \rotatebox{-90}{ZZXX} & \rotatebox{-90}{YZIX} & \rotatebox{-90}{XIXY} & \rotatebox{-90}{ZZXZ} & \rotatebox{-90}{YYXY} & \rotatebox{-90}{IXXX} & \rotatebox{-90}{YZIZ} & \rotatebox{-90}{Distance } 
    \\\hline (a) &1 & 1 & 1 & 1 & 0 & \textcircled{\textcolor{red}{2}} & 0 & 1 & 1 & 0 & \textcircled{\textcolor{red}{2}} & \textcolor{red}{2} & 0 & 1 & \textcircled{\textcolor{red}{2}} & \textbf{5}
    \\ (b) &0 & 1 & 1 & 0 & 0 & 1 & 0 & 0 & 1 & 0 & 1 & 1 & 0 & 1 & \textcircled{\textcolor{red}{2}} & \textbf{3}
    \\ (c) & 0 & 0 & 1 & 0 & 0 & 0 & 0 & 0 & 0 & 0 & 0 & 1 & 0 & 1 & 0 & \textbf{1}
\end{tabular}

\caption{Graphical illustration of the successive steps of the algorithm on a two-spread, showing the sign assigned to each observable and in the associated table the number of unsatisfied contexts (dashed lines) containing it. The negative contexts are represented by the doubled lines colored in red. The value $+1$ is first assigned to all 15 observables (a). The last step (c) reaches the minimal possible distance for this geometry. In this example, the threshold $\theta$ is $0.8$ and the sign flip probability $\gamma$ is $0.7$.}
\label{fig:algo-table}
\end{figure}

Once the whole assignment has been changed, the Hamming distance between it and the valuation vector is computed. If this distance is the smallest found thus far, the assignment is stored. This process continues until the number of iterations exceeds $\textit{MAX\_ITERATIONS}$. The purpose of this limit is to prevent an infinite loop.
Finally, this algorithm returns the best assignment $min\_a$ that was found, from which the corresponding unsatisfied configuration $\widetilde{{K}}^{{\tt uns}}$ can easily be computed. In the example in Figure~\ref{fig:algo-table}, since the algorithm starts by assigning to each observable the value $+1$, the only initially unsatisfied contexts are the five negative lines (dashed in  Figure~\ref{fig:algo-table}$(a)$). There are four observables featuring two, i.\,e. the maximal number of unsatisfied contexts passing through each of them, as also listed for the reader's convenience in the table at the bottom of the figure. In the next step, we flip the value from $+1$ to $-1$ at three of them, as indicated by circles in Figure~\ref{fig:algo-table}$(a)$ and the associated table as well. With this changed labeling we  find that at this step only three contexts are unsatisfied, as shown in Figure~\ref{fig:algo-table}$(b)$. As now there is only one observable, namely $YZIZ$, that is on two unsatisfied contexts, we just need one more step to get the final result, Figure~\ref{fig:algo-table}$(c)$, where only one context remains unsatisfied. It is worth noting that while every assignment given from a loop iteration in this example has a lower Hamming distance than the preceding one, this is not always the case generally.

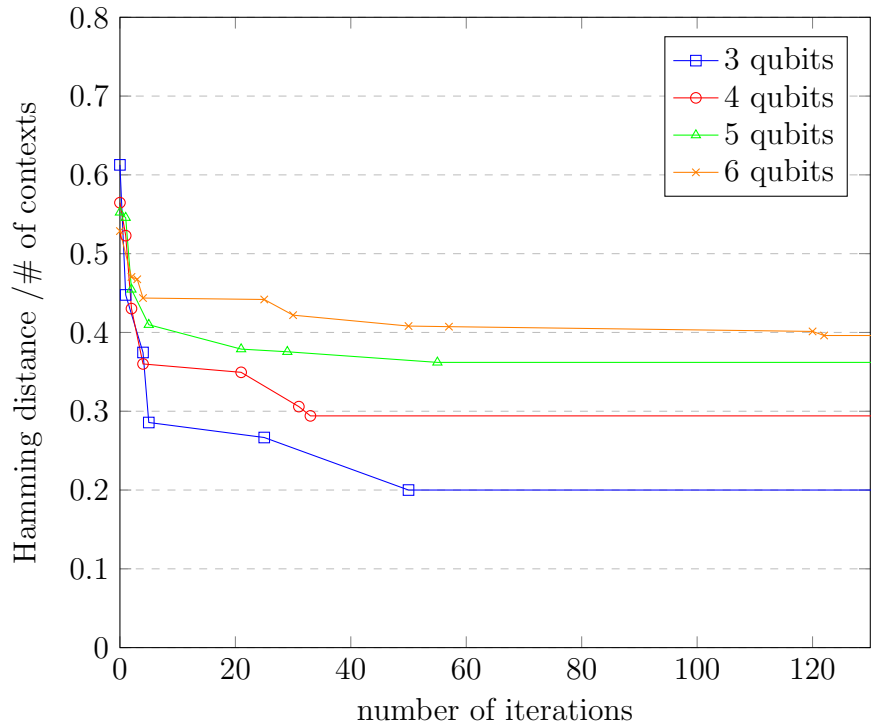
\begin{figure}[hbt!]
    \centering
    \begin{tikzpicture}
    \begin{axis}[
        xlabel={number of iterations}, ylabel={Hamming distance /\# of contexts},
        xmin=0, xmax=130,
        ymin=0, ymax=0.8,
        xtick={0,20,...,130},
        ytick={0,0.1,...,1},
        legend pos=north east,
        ymajorgrids=true,
        grid style=dashed,
        width=0.7\textwidth,
    ]
    
    \addplot[
        color=blue,
        mark=square,
        y filter/.expression={y/315},
        ]
        coordinates {
        (0,193) (1,141) (4,118) (5,90) (25,84) (50,63)      (1000,63)
        };
        \legend{$3$ qubits}

    \addplot[
        color=red,
        mark=o,
        y filter/.expression={y/5355},
        ]
        coordinates {
        (0,3024) (1,2800) (2,2304) (4,1928) (21,1871) (31,1638) (33,1575)      (1000,1575)
        };
        \addlegendentry{$4$ qubits}

    \addplot[
        color=green,
        mark=triangle,
        y filter/.expression={y/86955},
        ]
        coordinates {
        (0,48032) (1,47449) (2,39532) (5,35636) (21,32944) (29,32650) (55,31479)      (1000,31479)
        };
        \addlegendentry{$5$ qubits}

    \addplot[
        color=orange,
        mark=x,
        y filter/.expression={y/1396395},
        ]
        coordinates {
        (0,738219) (2,656892) (3,652791) (4,619539) (25,616998) (30,589224) (50,569892) (57,568824) (120,560384) (122,553140)     (1000,553140)
        
        };
        \addlegendentry{$6$ qubits}
    
    \end{axis}
    \end{tikzpicture}
    \caption{Minimal Hamming distances (the $y$-axis) per total number of contexts, over the number of iterations (the $x$-axis), computed by the heuristic method running simultaneously on 200 threads shared between 20 cores of an Intel(R) Core(TM) i7-12700H processor, for the quantum configurations composed of all the lines of the three- to six-qubit symplectic polar spaces.\label{HammingIterationsGraph}}
\end{figure}

\subsection{A brief inventory of new results}

This algorithm was run on the set of all lines of $\mathcal{W}(2N-1,2)$, $3 \leq N \leq 7$, as well as on one of its elliptic and hyperbolic quadrics, since all the other ones have the same degree of contextuality~\cites{mg24,TLC22}. The basic results are collected in Table~\ref{k1Table} for configurations comprising all the lines of the corresponding space and in Table~\ref{resultsTable1} (Appendix~\ref{quadricDegreeApp}) for quadrics. For each quantum geometry/configuration, the same algorithm was run on 200 threads, from which the best assignment was then selected. Various values were tested for $\theta$ and $\gamma$ (both between 0 and 1) simultaneously within each thread and, although some geometries have sometimes a converging assignment for specific values, it was found that with $\theta = 0.8$ and $\gamma = 0.9$ the algorithm generally gave quicker and better results. The overall performance of the algorithm is depicted in Figure~\ref{HammingIterationsGraph}, where only the best $min\_a$ assignment found at each iteration is plotted, hence the decreasing curves.

\begin{table}[bht!]
\begin{footnotesize}
\begin{center}
\begin{tabular}{|c|r|r|r|l|l|}
\hline
$N$   & $p$  & $l$          & $l^{-}$          & $d$  & Duration\\
\hline
\hline
2            & 15           & 15               & 3                & 3    & $<$ 1 s\\
\cline{1-6}
3            & 63           & 315              & 90               & $63$ & $<$ 1 s\\
\cline{1-6}
4            & 255          & \np{5355}        & \np{1908}        & $\leq\mathbf{\np{1575}}$ & $2$s\\
\cline{1-6}
5            & \np{1023}    & \np{86955}       & \np{35400}       & $\leq\mathbf{\np{31479}}$ & $<$ 1 mn\\
\cline{1-6}
6            & \np{4095}    & \np{1396395}     & \np{615888}      & $\leq\mathbf{\np{553140}}$ & $<$ 1 mn\\
\cline{1-6}
7            & \np{16383}    & \np{22362795}    & \np{10352160}    & $\leq\mathbf{\np{9406024}}$ & $<$ 10 mn\\
\hline
\end{tabular}
\end{center}
\end{footnotesize}
\caption{Exact values or specific upper bounds for the contextuality degree $d$ of
quantum configurations isomorphic to the configuration whose contexts
are all the lines of $\mathcal{W}(2N-1,2)$, for $3 \leq N \leq 7$. Here,
$p$ is the number of observables, $l$ is the total number of contexts and $l^{-}$  is the number of negative contexts in a given configuration.
The improved upper bounds are highlighted in the \textbf{bold font}. \label{k1Table}}
\end{table}

\section{Combinatorial geometric estimates of some specific lower bounds}\label{sec:lobo}

Let $d^{\rm full}_N$ be the degree of contextuality for the configuration comprising all the lines/con\-texts in $\mathcal{W}(2N-1,2)$. Generalizing to an arbitrary $N \geq 3$ the chain of group-geometrical arguments we and Henri de Boutray employed in a previous work~\cite[Sec.\,4.1]{Sanigaetal23}, we have the following recurrent formula for these degrees

\[
d^{\rm full}_N \geq \frac{\# {\rm ~of~ quadratic~} \mathcal{W}(2N-3,2)'s~ {\rm in~} \mathcal{W}(2N-1,2)}{\# {\rm ~of~ quadratic~} \mathcal{W}(2N-3,2)'s~ {\rm on~a~line~ in~} \mathcal{W}(2N-1,2)} d^{\rm full}_{N-1},
\]
which, using the combinatorial properties of symplectic polar spaces, can be cast into a simpler form,
\[
d^{\rm full}_N \geq \frac{\# {\rm ~of~ points~ in~} \mathcal{W}(2N-1,2)}{\# {\rm ~of~ points~ in~} \mathcal{W}(2N-5,2)} d^{\rm full}_{N-1},
\]
which explicitly reads (see eq. (\ref{eq:w-p}))
\begin{equation}
d^{\rm full}_N \geq \frac{4^N -1}{4^{N-2} -1} d^{\rm full}_{N-1}.
\label{recurr}
\end{equation}
Employing the well-established fact that $d^{\rm full}_2 = 3$, we get
\begin{equation}
d^{\rm full}_N \geq \frac{\left(4^N -1\right)\left(4^{N-1} -1\right)}{15}.
\label{recurr}
\end{equation}
The values of this lower bound up to nine qubits are as listed in Table~\ref{tab-lb}.

\begin{table}[H]
\begin{center}
\begin{tabular}{cc}
\hline
$N$ &  $d^{\rm full}_N \geq$  \\
\hline
 2 & 3    \\
 3 & 63    \\
 4 & \np{1071}   \\
 5 & \np{17391}   \\
 6 & \np{279279}   \\
 7 & \np{4472559}   \\
 8 & \np{71577327}   \\
 9 & \np{1145302767}   \\
 \hline
\end{tabular}
\end{center}
\caption{Lower bound for the degree of contextuality of contextual configurations comprising all the lines of $\mathcal{W}(2N-1,2)$ of small rank.}
\label{tab-lb}
\end{table}

Another set of sufficient (but not necessary) criteria for ascertaining the former lower bound on the contextuality degree of the whole $\mathcal{W}(2N-1,2)$, with $N > 2$, is as follows:
\begin{equation}
{F}^{{\tt uns}}_{N} \cap \mathcal{W}(2N-3,2) \cong {F}^{{\tt uns}}_{N-1}
{\rm ~for~any~} \mathcal{W}(2N-3,2) \in \mathcal{W}(2N-1,2),
\label{cr-full}
\end{equation}
\begin{equation}
{F}^{{\tt uns}}_{N} \cap \mathcal{Q}^{-}(2N-1,2) \cong {E}^{{\tt uns}}_{N}
{\rm ~for~any~} \mathcal{Q}^{-}(2N-1,2) \in  \mathcal{W}(2N-1,2)
\label{cr-ell}
\end{equation}
and 
\begin{equation}
{F}^{{\tt uns}}_{N} \cap \mathcal{Q}^{+}(2N-1,2) \cong {H}^{{\tt uns}}_{N}
{\rm ~for~any~} \mathcal{Q}^{+}(2N-1,2) \in  \mathcal{W}(2N-1,2).
\label{cr-hyp}
\end{equation}
In the case of $N=3$, with the minimal unsatisfied configuration ${F}^{{\tt uns}}_{N} \cong \mathcal{H}_C$ and so with  $d^{\rm full}_3$ reaching its lower bound 63, from the analysis carried out in~\cite{Sanigaetal23} it follows that all the three criteria are here indeed satisfied. In the next section we will see that this is not the case for $N=4$.

\section{Geometries and graphs underpinning the most illustrative four- to six-qubit examples}\label{sec:res}

As it has already been stressed, with the code based on a SAT solver
we were unable to properly address contextuality issues for symplectic polar spaces
of rank greater than three. To illustrate the power of our new approach, we
will discuss in detail the first open case in the hierarchy, $N=4$, and then briefly address also the $N=5$ and $N=6$ cases.

\subsection{Contextuality in the four-qubit space}

\subsubsection{Contextuality of elliptic quadrics}

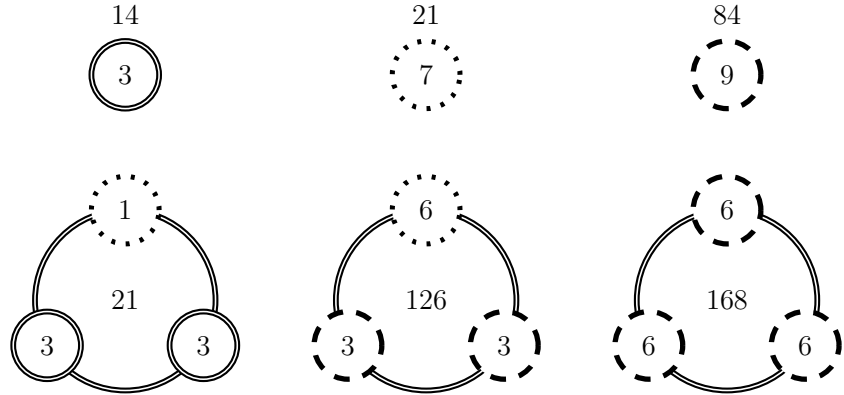
\begin{figure}[pth!]

\begin{center}
\begin{tikzpicture}[every plot/.style={smooth, tension=2},
  scale=1, 
  every node/.style={scale=0.9,fill=white,draw,circle,line width = 0.7mm,minimum size=1cm}
]
    \def\radius{1.2cm}
    \node[draw=none] at (0,3.8) {14}; 
    \node[double,thick] at (0,3) {3};

    \draw[double,thick] (0,0) circle (\radius);  
    \node[draw=none] at (0,0) {21}; 
    \node[dottednode] at (90:\radius) {1};
    \node[double,thick] at (210:\radius) {3};
    \node[double,thick] at (330:\radius) {3};
    
    \begin{scope}[shift={(4cm,0)}] 
        \node[draw=none] at (0,3.8) {21}; 
        \node[dottednode] at (0,3) {7};
    
        \draw[double,thick] (0,0) circle (\radius);  
        \node[draw=none] at (0,0) {126}; 
        \node[dottednode] at (90:\radius) {6};
        \node[dashednode] at (210:\radius) {3};
        \node[dashednode] at (330:\radius) {3};
    \end{scope}
    \begin{scope}[shift={(8cm,0)}] 
        \node[draw=none] at (0,3.8) {84}; 
        \node[dashednode] at (0,3) {9};
    
        \draw[double,thick] (0,0) circle (\radius);  
        \node[draw=none] at (0,0) {168}; 
        \node[dashednode] at (90:\radius) {6};
        \node[dashednode] at (210:\radius) {6};
        \node[dashednode] at (330:\radius) {6};
    \end{scope}
\end{tikzpicture}
\end{center}

\caption{Properties of the point-line geometry comprising 315 unsatisfied
constraints of a particular elliptic quadric whose index is $IIIY$. For a point on a line, the number inside the circle
corresponds to its restricted degree in the configuration consisting solely of lines of this particular type.}
\label{4q-pattern-ell}
\end{figure}

A four-qubit elliptic quadric, $\mathcal{Q}^{-}(7,2)$, when viewed as a point-line incidence
structure, features 119 points and \np{1071} lines, with 27 lines on a point and three points
on a line. The smallest number of negative lines a $\mathcal{Q}^{-}(7,2)$ can have
is 360, which is thus a first upper bound on its contextuality degree $d^{\rm ell}_4$, i.\,e. $d^{\rm ell}_4 \leq 360$. 
Using our new approach, we were able to push this value much lower, namely to $d^{\rm ell}_4 \leq 315$ (see Table~\ref{resultsTable1} in Appendix~\ref{quadricDegreeApp}).
The found configuration of 315 unsatisfied contexts, ${E}^{{\tt uns}}_{4}$, encompassing all the points of the quadric, 
features 14 points of degree three (let us call them solids), 21 points of degree seven (dots) and 
84 points of degree nine (dashes), as also illustrated in a graphical form in Figure~\ref{4q-pattern-ell}, top layer. Out of its 315 lines, there are 21 of type solid-solid-dot, 126 of type
dash-dash-dot and, finally, 168 of type dash-dash-dash, as illustrated in Figure~\ref{4q-pattern-ell},
bottom layer.

\begin{figure}[pth!]
\centerline{\includegraphics[width=4.0truecm,clip=]{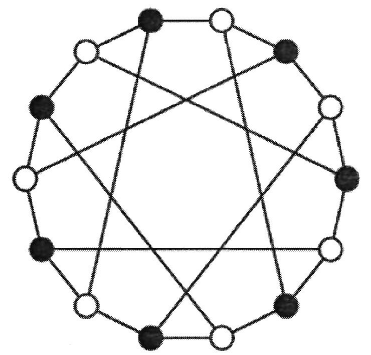} \hspace{3cm}\includegraphics[width=4.0truecm,clip=]{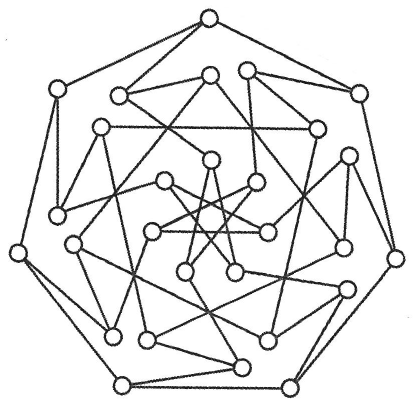}}
\vspace*{.2cm}
\caption{An illustration of the Heawood graph (left) and the Coxeter one (right), both drawn with seven-fold rotational symmetry.}
\label{heco}
\end{figure}

In each line of the first type, the restricted degree of a solid point is three and that of a dotted one is of one.
Moreover, the 21 lines of this form can uniquely be associated with the edges of the well-known Heawood graph~\cite{heaw} -- see Figure~\ref{heco}, left -- in such a way that
the 14 solid points will be its vertices and the 21 dotted points will be the third points on its edges so that these edges
become lines of $\mathcal{W}(7,2)$; hence, this 21-line configuration is isomorphic to nothing but the configuration representing
21 unsatisfied contexts of a three-qubit hyperbolic quadric~\cite{muller2023new}.

\begin{figure}[t]
\centerline{\includegraphics[width=9.0truecm,clip=]{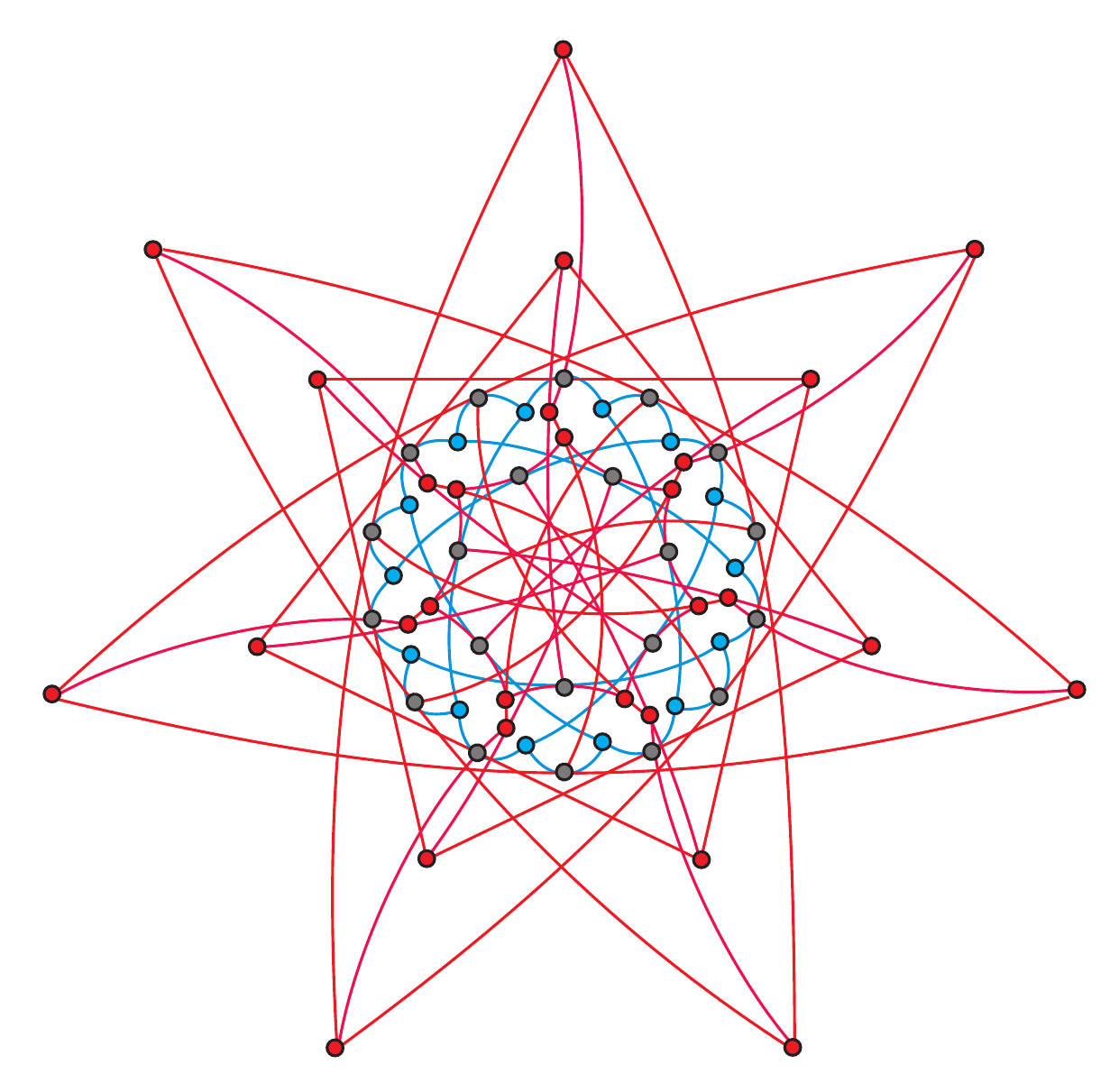}}
\vspace*{.2cm}
\caption{A generic layering of each of the three split Cayley hexagons of order two encapsulating the core of ${E}^{{\tt uns}}_{4}$.
The 14 solid points are colored blue, the 21 dotted points are gray and 28 out of 84 dashed points are represented by red color. The 21 blue lines are the Heawood lines, those colored red are the Coxeter ones. Removing from the hexagon the 21 gray points we indeed get a disjoint union of the Heawood graph and the Coxeter graph.}
\label{hexagon}
\end{figure}

Next, let us consider the second configuration of lines.
In each of these 126 lines, the restricted degree of a dashed point is three whilst that of a dotted one is six. These 126 lines split into three disjoint, equally-sized sets that are isomorphic to each other. Given such a set, if one considers its 28 dashed points as vertices and the corresponding 42 lines as edges, one obtain a graph that, remarkably, is isomorphic to the famous Coxeter graph~\cite{cox} -- depicted in Figure~\ref{heco}, right. Moreover, joining such a set of 42 `Coxeter' lines\footnote{It is worth mentioning here that the subgeometry of the split Cayley hexagon of order two related to (or underpinned by) its Coxeter graphs was some 15 years ago found to be intricately related to the $E_7$-symmetric black-hole entropy formula in string theory~\cite{lsv}.} with the 21 `Heawood' lines of the first type we get nothing but a copy of the split Cayley hexagon of order two, as schematically depicted in Figure~\ref{hexagon} (this being a simplified reproduction of part of Figure 7 of~\cite{psm}; see also~\cites{schr,polmal}). 
Each of the three split Cayley  hexagons is classically embedded into a parabolic quadric $\mathcal{Q}(6,2)$ in which our elliptic quadric cuts a certain PG$(6,2)$ of the ambient space PG$(7,2)$; the three corresponding PG$(6,2)$s have a PG$(5,2)$ in common, the latter cutting our 
$\mathcal{Q}^{-}(7,2)$ in a hyperbolic quadric, $\mathcal{Q}^{+}(5,2)$ -- the one accommodating the 14 points of degree three and the 21 points of degree seven.
It is worth mentioning here that the 21 points of degree seven form in each hexagon a
geometric hyperplane; this smallest-size hyperplane is called a distance-2 ovoid and is of type
$\mathcal{V}_2$ in the notation of Frohardt and Johnson~\cite{fj}.

Finally, each of the 168 lines of the last type is such that it shares a single point with each of the three hexagons. Hence, the 12 points located on six lines passing via a dashed point split into two sextuples, either sextuple being located in a hexagon. In addition, each sextuple further splits into a pair of tricentric triads that define a unique quadratic doily.
\subsubsection{Contextuality of hyperbolic quadrics}

A four-qubit hyperbolic quadric, $\mathcal{Q}^{+}(7,2)$, when viewed as a point-line incidence
structure, features 135 points and \np{1575} lines, with 35 lines on a point and three points
on a line. The smallest number of negative lines a $\mathcal{Q}^{+}(7,2)$ can have
is 532, hence its degree of contextuality $d^{\rm hyp}_4$ satisfies $d^{\rm hyp}_4 \leq 532$. As in the preceding case, we can do much better with our new approach, namely
$d^{\rm hyp}_4 \leq 315$ (see Table~\ref{resultsTable1} in Appendix~\ref{quadricDegreeApp}). Here, the pattern of 315 unsatisfied contexts we have discovered, ${H}^{{\tt uns}}_{4}$, that again covers all the points of the quadric,  is much more symmetric than that characterizing an elliptic quadric, since through each point there pass the same number of lines, namely seven. Using the Lagrangian Grassmannian mapping of the type $LGr(3,6)$ (see, e.\,g., \cites{lps13,hsl14}) that sends planes of 
$\mathcal{W}(5,2)$ into points of a certain $\mathcal{Q}^{+}(7,2)$ and lines of 
$\mathcal{W}(5,2)$ into lines of the same quadric (and whose explicit form we made use of is given in Appendix~\ref{app}),  this $(135_7, 315_3)$-configuration is found to be, in fact, 
isomorphic to $\mathcal{DW}(5,2)$. This is a very important fact in the following sense. We know that each
$\mathcal{Q}^{+}(7,2)$ contains 120 quadratic $\mathcal{W}(5,2)$s. On the other hand, $\mathcal{DW}(5,2)$ contains
the same number of copies of the split Cayley hexagon of order two (these being, in fact, its geometric hyperplanes, see, e.\,g., \cite{DeBruyn}). Hence,   
when a $\mathcal{DW}(5,2)$ is embedded into a $\mathcal{Q}^{+}(7,2)$, each hexagon must be hosted by a unique $\mathcal{W}(5,2)$.
In other words, our unsatisfied configuration ${H}^{{\tt uns}}_{4}$ picks up from (or shares with) {\it each} $\mathcal{W}(5,2)$ of the $\mathcal{Q}^{+}(7,2)$ a single copy of the split Cayley hexagon of order two; moreover, we have verified that it is always a copy that is {\it classically}-embedded into the corresponding $\mathcal{W}(5,2)$. Given the fact that any unsatisfied configuration of the full three-qubit $\mathcal{W}(5,2)$ is isomorphic to
a copy of the smallest split Cayley hexagon embedded classically into the space, ${F}^{{\tt uns}}_{3} \cong \mathcal{H_C}$~\cites{muller2023new, Sanigaetal23}, and a very recent proof~\cite{mg24} that this isomorphism must hold for any $\mathcal{W}(5,2)$ located in any higher-rank space,  our unsatisfied ${H}^{{\tt uns}}_{4} \cong \mathcal{DW}(5,2)$ behaves exactly as one would expect for a configuration that also gives the {\it lower} bound for the contextuality degree of the four-qubit $\mathcal{Q}^{+}(7,2)$. Moreover, we have also verified that both
$\mathcal{DW}(5,2)$ and its complement on the $\mathcal{Q}^{+}(7,2)$ are, like an $\mathcal{H_C}$ and its complement in $\mathcal{W}(5,2)$, {\it not} contextual.
There is also a neat combinatorial argument speaking in favor of our conjecture.
There are altogether 136 $\mathcal{Q}^{+}(7,2)$'s in $\mathcal{W}(7,2)$, each of them has \np{1575} lines and as there are \np{5355} lines in $\mathcal{W}(7,2)$, each line of this space is shared by $136 \times 1575/5355 = 40$  $\mathcal{Q}^{+}(7,2)$'s. And, remarkably,
$136 \times 315 / 40 = \np{1071}$, which is indeed equal to the lower bound ascertained in Section~\ref{sec:lobo} (see Table~\ref{tab-lb}) for the degree of contextuality of the configuration comprising all the lines of $\mathcal{W}(7,2)$.
So, we do believe that $d^{\rm hyp}_4 = 315$, with the understanding that the corresponding unsatisfied configurations are (always) isomorphic to $\mathcal{DW}(5,2)$.

\subsubsection{Contextuality of the full four-qubit space}

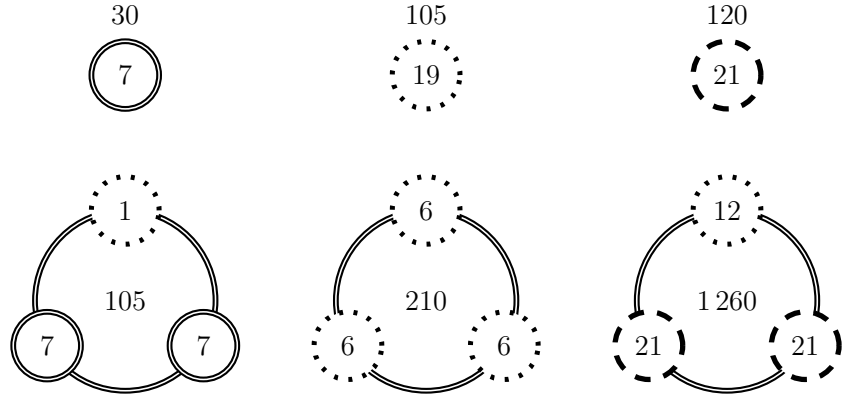
\begin{figure}[pth!]

\begin{center}
\begin{tikzpicture}[every plot/.style={smooth, tension=2},
  scale=1, 
  every node/.style={scale=0.9,fill=white,draw,circle,line width = 0.7mm,minimum size=1cm}
]
    \def\radius{1.2cm}
    \node[draw=none] at (0,3.8) {30}; 
    \node[double,thick] at (0,3) {7};
    
    \draw[double,thick] (0,0) circle (\radius);  
    \node[draw=none] at (0,0) {105}; 
    \node[dottednode] at (90:\radius) {1};
    \node[double,thick] at (210:\radius) {7};
    \node[double,thick] at (330:\radius) {7};
    
    \begin{scope}[shift={(4cm,0)}] 
        \node[draw=none] at (0,3.8) {105}; 
        \node[dottednode] at (0,3) {19};
    
        \draw[double,thick] (0,0) circle (\radius);  
        \node[draw=none] at (0,0) {210}; 
        \node[dottednode] at (90:\radius) {6};
        \node[dottednode] at (210:\radius) {6};
        \node[dottednode] at (330:\radius) {6};
    \end{scope}
    \begin{scope}[shift={(8cm,0)}] 
        \node[draw=none] at (0,3.8) {120}; 
        \node[dashednode] at (0,3) {21};
    
        \draw[double,thick] (0,0) circle (\radius);  
        \node[draw=none] at (0,0) {\np{1260}}; 
        \node[dottednode] at (90:\radius) {12};
        \node[dashednode] at (210:\radius) {21};
        \node[dashednode] at (330:\radius) {21};
    \end{scope}
\end{tikzpicture}
\end{center}

\caption{Properties of the point-line geometry comprising \np{1575} unsatisfied
constraints for the contextual geometry whose contexts are all the lines
of the space $\mathcal{W}(7,2)$.}
\label{4q-pattern-full}
\end{figure}

The full four-qubit space, $\mathcal{W}(7,2)$, features 255 points and \np{5355} lines, with 63 lines through a point  and three points on a line. As \np{1908} out of its \np{5355} contexts are negative, 
$d^{\rm full}_4 \leq 1908$. Also this bound has been considerably reduced, down to \np{1575} (see Table~\ref{k1Table}). This bound is also the number of lines on a $\mathcal{Q}^{+}(7,2)$ -- an intriguing coincidence.
The corresponding configuration of \np{1575} unsatisfied contexts, ${F}^{{\tt uns}}_{4}$, encompassing all the 255 points of the space, 
features 30 points of degree seven (to be referred to as solids), 105 points of degree 19 (dots) and 
120 points of degree 21 (dashes), as also illustrated in a graphical form in Figure 
\ref{4q-pattern-full}, top layer. Out of its \np{1575} lines, there are 105 of type solid-solid-dot, 210 of type
dot-dot-dot and, finally, \np{1260} of type dash-dash-dot, as illustrated in Figure~\ref{4q-pattern-full}, bottom layer. 

\begin{figure}[t]
\centerline{\includegraphics[width=12.5truecm,clip=]{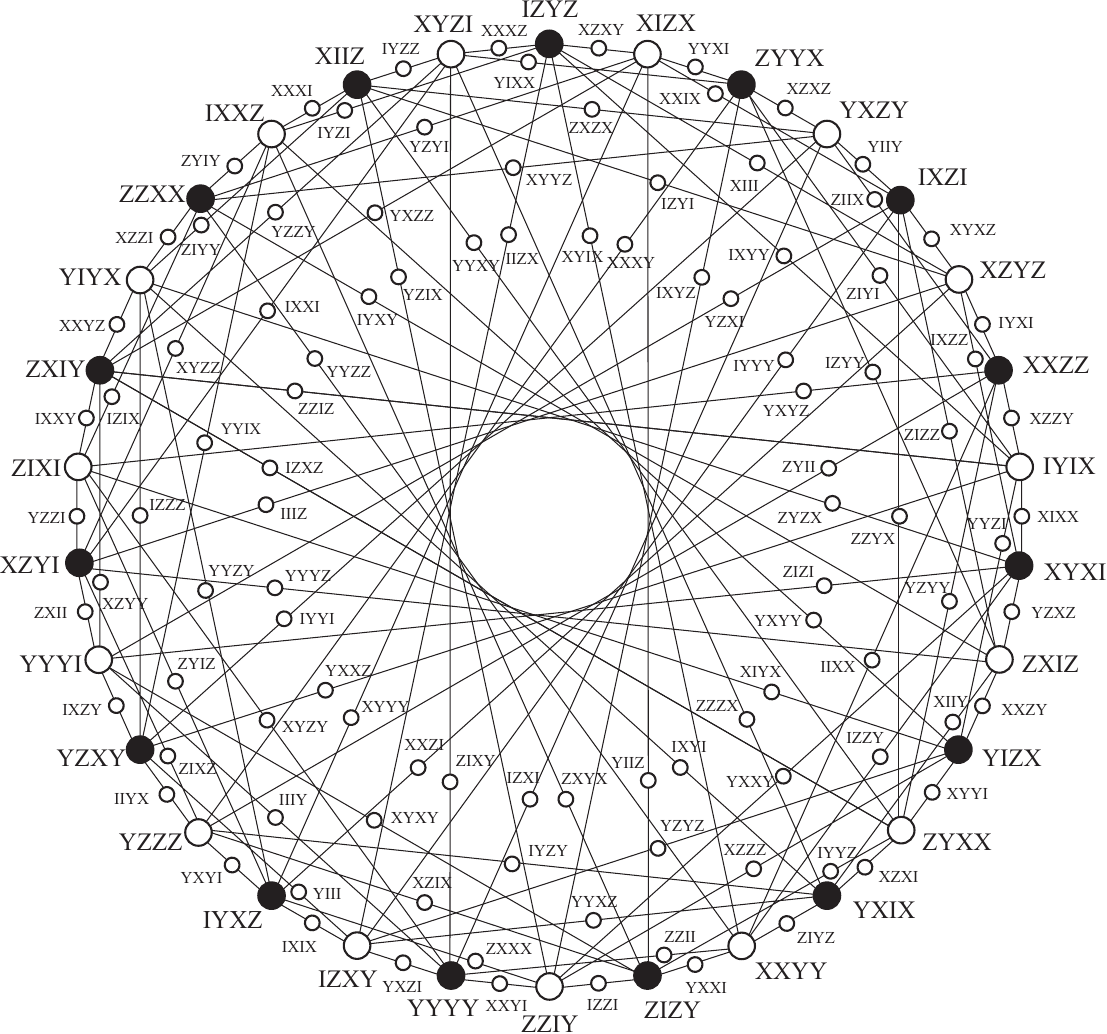}}
\vspace*{.2cm}
\caption{A graphical representation of the subgeometry formed by the 105 lines of the first type. The underlying point-plane incidence graph of PG$(3,2)$ -- whose vertices are represented by 15 big white circles as well as 15 big black circles (bullets) -- is rendered in the rotationally symmetric form isomorphic to the Haar graph $H(17051)$. }
\label{4q-ppig-pg32}
\end{figure}

In each line of the first type, the restricted degree of a solid point is seven and that of a dotted one is of one.
The 105 lines of this form can uniquely be associated with the edges of the point-plane incidence graph of PG$(3,2)$ in such a way that
the 30 solid points will be its vertices and the 105 dotted points will be the third points on its edges so that these edges
become lines of $\mathcal{W}(7,2)$ -- as also illustrated in Figure~\ref{4q-ppig-pg32}. Moreover, these lines together with the 210
lines of the second type form a geometry isomorphic to $\mathcal{DW}(5,2)$, which is fully located on a particular hyperbolic quadric that consists of the 30 points of degree seven
and the 105 points of order 19 (and whose index, as readily discernible from Figure~\ref{4q-ppig-pg32}, is $XYYZ$). Interestingly, the 105 points of order 19 form in the $\mathcal{DW}(5,2)$ a geometric hyperplane, namely the one belonging to class 3 in the
classification of Pralle~\cite{pralle}.

\begin{figure}[h]
\centerline{\includegraphics[width=14.0truecm,clip=]{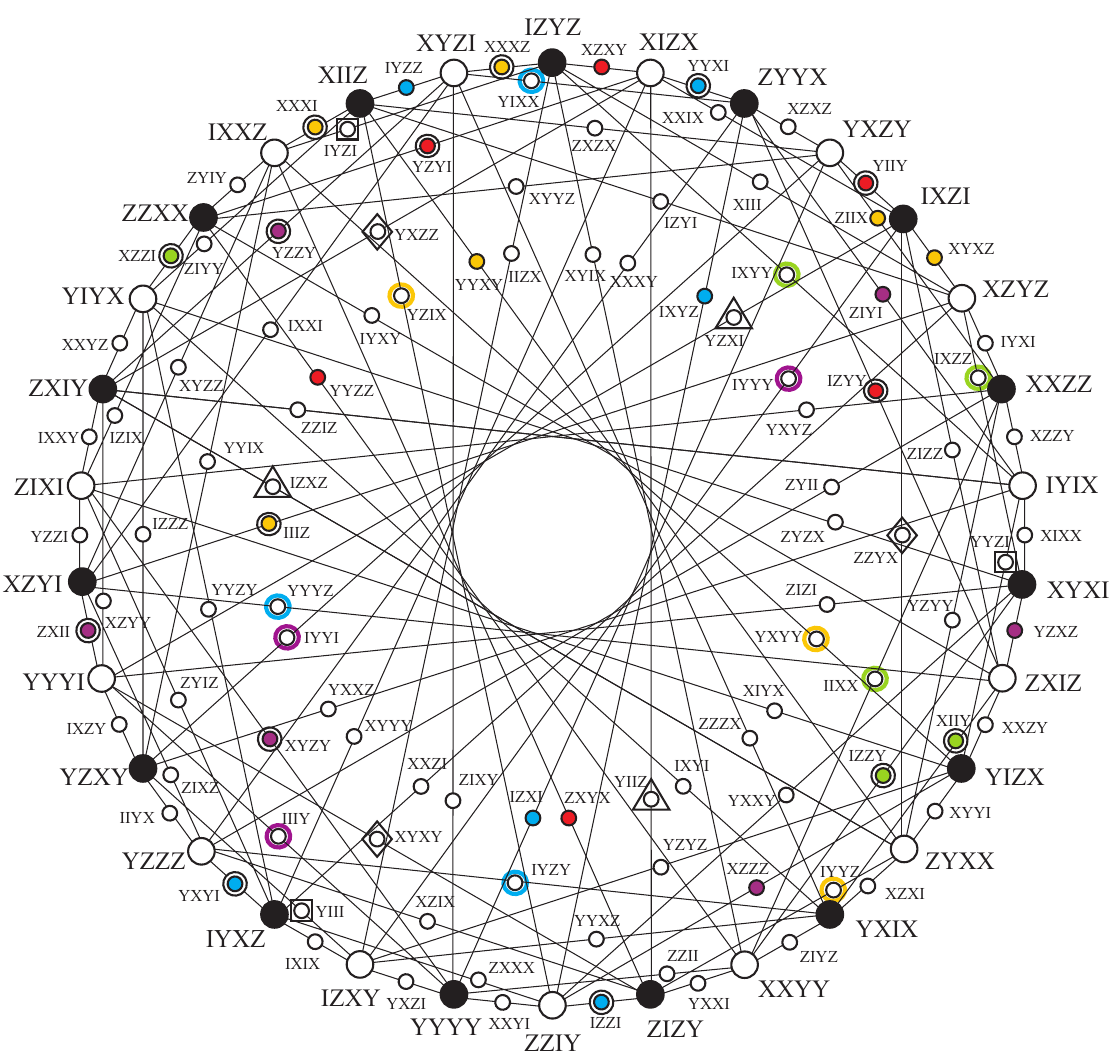}}
\vspace*{.2cm}
\caption{An illustration of 16 out of the 210 lines of type two. The remaining lines from each orbit can be obtained by successive rotations of the figure through 360/15 degrees around its center while keeping the position of each label/observable fixed.}
\label{DW52-210}
\end{figure}

Using our Haar-graph-based geometry of   Figure~\ref{4q-ppig-pg32}, the 210 lines of the second type are found to form 13 orbits of size 15  and three orbits of size five with respect to the action of the automorphism of order 15 of the figure.  A representative line for each of these orbits is portrayed  in Figure~\ref{DW52-210}.
In particular, we have for the encircled-filled part the lines $YYXI-YXYI-IZZI$ (blue),
$XZZI-IZZY-XIIY$ (green), $YZYI-IZYY-YIIY$ (red), $XXXZ-XXXI-IIIZ$ (yellow) and
$YZZY-ZXII-XYZY$ (violet); for the plain-filled family the lines $IYZZ-IXYZ-IZXI$ (blue), $XZXY-ZXYX-YYZZ$ (red), $YYXY-ZIIX-XYXZ$ (yellow) and $ZIYI-YZXZ-XZZZ$ (violet); and for
the circled color set the lines $YIXX-YYYZ-IYZY$ (blue), $IXYY-IXZZ-IIXX$ (green),
$YZIX-YXYY-IYYZ$ (yellow) and $IYYI-IIIY-IYYY$ (violet). The three representatives
of size-five orbits are: $IYZI-YIII-YYZI$ (squares),  $YXZZ-XYXY-ZZYX$ (diamonds) and
$IZXZ-YIIZ-YZXI$ (triangles).

\begin{figure}[h]
\centerline{\includegraphics[width=14.0truecm,clip=]{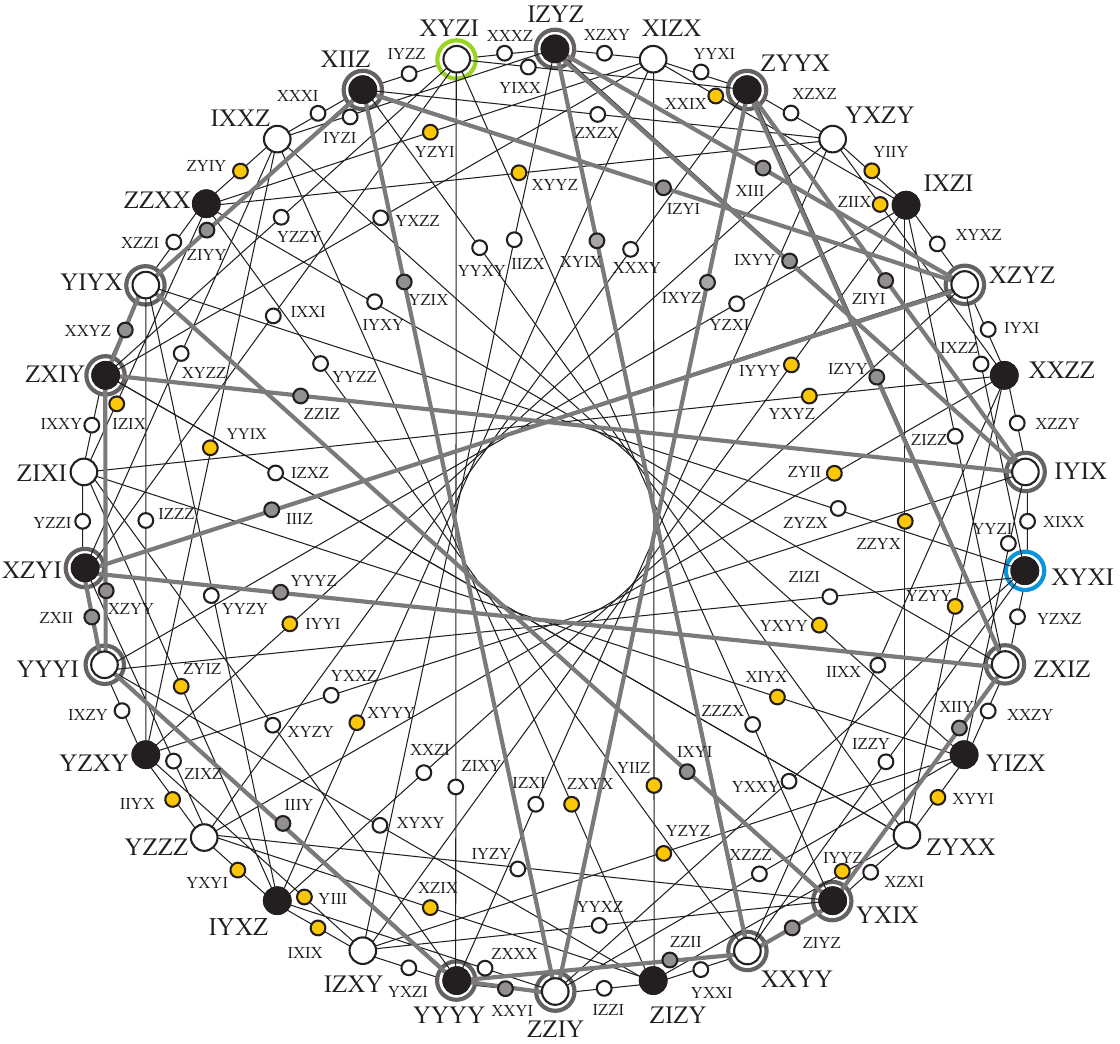}}
\caption{An illustration of main properties of the points and lines of the third type. The meaning of the
colored points and highlighted lines is described in the text.}
\label{21pts}
\end{figure}

Let us, finally, focus on the dash-type points and the lines of the last, i.e. dash-dash-dot type.
The 120 dash-type points are exactly those points that lie out off the
hyperbolic quadric $\mathcal{Q}^{+}_{XYYZ}(7,2)$ that hosts our unsatisfied $\mathcal{DW}(5,2)$.
Through each of these points there pass 21 lines of the above-mentioned type that cut
the $\mathcal{Q}^{+}_{XYYZ}(7,2)$ in the 21 points of the dot-type. By a way of example,
the 21 lines through the (off-quadric) point $IIYI$ cut the $\mathcal{Q}^{+}_{XYYZ}(7,2)$ in the 21 points
shown in Figure~\ref{21pts} in gray color. Each of these points is incident with a unique line  of the first type (bold gray) whose
other two points are the vertices of the point-plane incidence graph of PG$(3,2)$, illustrated in 
Figure~\ref{21pts} by big gray circles. There are 14 such vertices that together with 21 distinguished edges
form a subgraph of the point-plane incidence graph of PG$(3,2)$ that is isomorphic to the Heawood graph.
The 35 points of this Heawood-graph-underpinned geometry define a certain $\mathcal{\widetilde{Q}}^{+}(5,2)$ in some PG$(5,2)$ of the ambient PG$(7,2)$. Our selected point $IIYI$ is also the nucleus of a unique quadratic
$\mathcal{\widetilde{W}}(5,2) \in \mathcal{Q}^{+}_{XYYZ}(7,2)$ that contains $\mathcal{\widetilde{Q}}^{+}(5,2)$; the 28 points 
of  $\mathcal{\widetilde{W}}(5,2)$ that lie off $\mathcal{\widetilde{Q}}^{+}(5,2)$ are show in Figure~\ref{21pts} by yellow
color. Next, there exists a unique elliptic quadric, $\mathcal{\widetilde{Q}}^{-}(7,2) \in \mathcal{W}(7,2)$ that shares with
$\mathcal{Q}^{+}_{XYYZ}(7,2)$ our $\mathcal{\widetilde{W}}(5,2)$. Its remaining $(119 - 63 =)$ 56 points can be found as follows.
We consider the (non-isotropic) line that is polar to the above-defined PG$(5,2)$ with respect to the symplectic polarity defining $\mathcal{W}(7,2)$. This line passes, obviously, through the point $IIYI$
and its remaining two points are found among the vertices of our point-plane incidence graph of PG$(3,2)$
-- as also illustrated in Figure~\ref{21pts}. One of them is
$XYZI$ (green), which is exactly the vertex connected to those seven vertices of the Heawood (sub-)graph that are represented by bullets, whereas the other one, $XYXI$ (blue), is that vertex that is connected to the other set
of seven vertices, represented by big circles. As neither of the two points lies on the $\mathcal{\widetilde{Q}}^{-}(7,2)$,
they are the nuclei of two different quadratic $\mathcal{W}(5,2)'$ and $\mathcal{W}(5,2)''$ lying on $\mathcal{\widetilde{Q}}^{-}(7,2)$ and having $\mathcal{\widetilde{Q}}^{+}(5,2)$ in common; our remaining 56 points of $\mathcal{\widetilde{Q}}^{-}(7,2)$
are nothing but those 56 points (of dash type) that belong to the symmetric difference of  $\mathcal{W}(5,2)'$ and $\mathcal{W}(5,2)''$. It is a straightforward, though by hand a bit lengthy, task to verify that ${F}^{{\tt uns}}_{4}$ shares with both $\mathcal{W}(5,2)'$ and $\mathcal{W}(5,2)''$, like with
$\mathcal{\widetilde{W}}(5,2)$ itself, a (classically-embedded) copy of the split Cayley hexagon of order two, the three hexagons having the above-described Heawood-graph-underpinned configuration in common. And this remarkable property holds if we take instead $IIYI$ any other point of the dash type.

\begin{figure}[t]
\centerline{\includegraphics[width=14.0truecm,clip=]{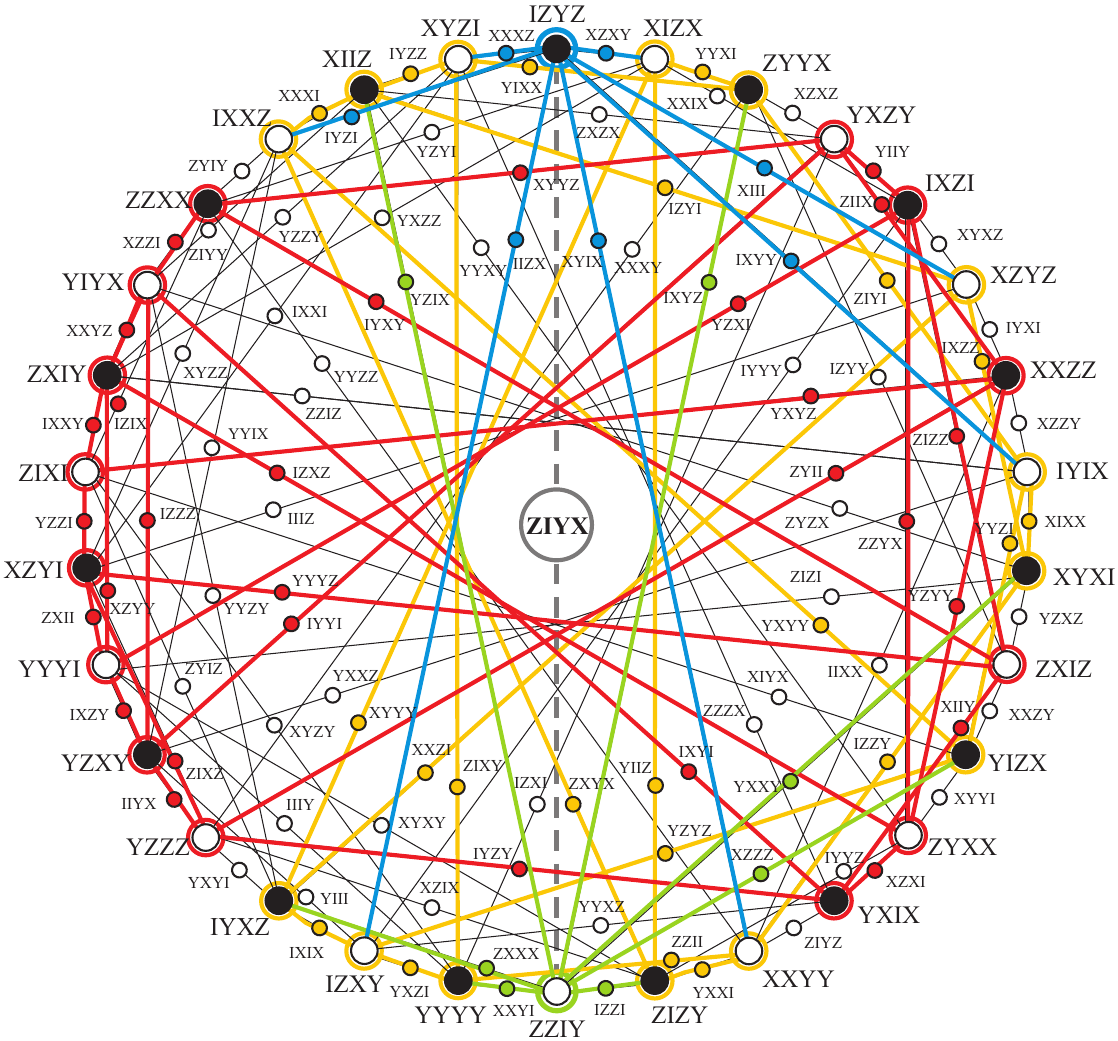}}
\caption{An illustration of the reversed procedure. Here, $\mathcal{O}_f = IZYZ$,
$\mathcal{O}_e = ZZIY$ and $\mathcal{O}_f . \mathcal{O}_e = ZIYX$. The seven neighbors
of $\mathcal{O}_f$ are pointed out by blue edges, those of $\mathcal{O}_e$  by green ones and the associated Heawood graph is highlighted in yellow color. (Interestingly, the complement of the Heawood graph, that is the quartic bipartite graph on 14 vertices and 28 edges highlighted in red color, is nothing but the Levi graph of the biplane of order two.)}
\label{4q-to-uns-full}
\end{figure}

Obviously, reversing the above-given chain of reasoning helps us find all \np{1260} lines of the third type solely from the configuration formed by the 105 type-one lines, once the latter are represented as portrayed in Figure~\ref{4q-ppig-pg32}. A specific case of this reversed procedure is portrayed in Figure~\ref{4q-to-uns-full}.
 Let us start with Figure~\ref{4q-ppig-pg32}. In its underlying point-plane incidence graph of PG$(3,2)$, let us pick up an arbitrary black vertex, $\mathcal{O}_f$, and an arbitrary white vertex not adjacent to it, $\mathcal{O}_e$; the third point on the line
defined by the two vertices, $\mathcal{O}_f . \mathcal{O}_e$, lies clearly  off the quadric $\mathcal{Q}^{+}_{XYYZ}(7,2)$ because the two corresponding observables anti-commute (and so the line does not belong to $\mathcal{W}(7,2)$ and, hence, to $\mathcal{Q}^{+}_{XYYZ}(7,2)$). Next, take all the seven white vertices adjacent to
$\mathcal{O}_f$ as well as all the seven black ones adjacent to $\mathcal{O}_e$; these 14 vertices together with the corresponding 21 edges inherited from the
point-plane incidence graph of PG$(3,2)$ form a graph that is isomorphic to the Heawood graph.
Connecting the point
$\mathcal{O}_f . \mathcal{O}_e$ with  each of the 21 mid-points situated  on the edges of this Heawood graph one gets all the 21 lines of type three passing via this particular
off-quadric point. Now, since there are $15 \times (15-7) = 120$ black-white vertex pairs of the above-defined type and  no two such pairs define the same off-quadric point (otherwise the corresponding four vertices would be coplanar, which is impossible), this construction yields indeed all $120 \times 21/2= 1260$ lines of the  dash-dash-dot type.

\begin{figure}[t]
\centerline{\includegraphics[width=9.0truecm,clip=]{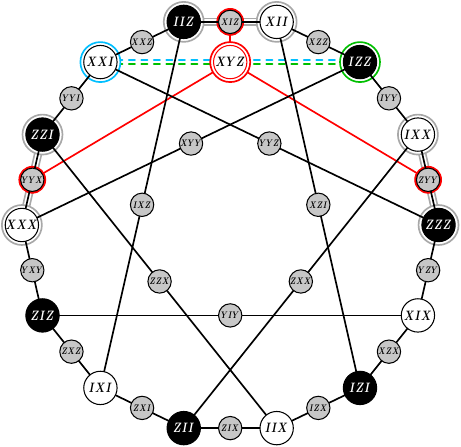}}
\caption{An illustration showing how, starting with the Heawood-graph-underpinned configuration in the three-qubit polar space, $\mathcal{W}(5,2)$, we can get the remaining 28 points and 42 lines of a classically-embedded copy of the split Cayley hexagon of order two. The role of color-highlighted elements is explained in the main text.}
\label{3q-to-uns-full}
\end{figure}

At this point, it is particularly instructive to make a slight digression from the main course of the paper and show that the just-outlined reversed procedure
yields in the three-qubit case a copy of the split Cayley hexagon of order two, i.\,e.
${F}^{{\tt uns}}_{3}$, and in the two-qubit doily just three pairwise disjoint lines, i.\,e. ${F}^{{\tt uns}}_{2}$. The three-qubit analogue of $\mathcal{DW}(5,2)$ is a Heawood-graph-underpinned configuration~\cites{muller2023new,Sanigaetal23} featuring 35 points and 21 lines, which can be taken -- without any substantial loss of generality -- to be located on the symmetric hyperbolic quadric, $\mathcal{Q}^{+}_{III}(5,2)$, as portrayed in Figure~\ref{3q-to-uns-full}.
From the underlying Heawood graph we pick up an arbitrary white vertex, say $XXI$ (blue), and any black one that is not adjacent to it, say $IZZ$ (green). The two vertices define a non-isotropic line of the ambient PG$(5,2)$ (dashed blue-green) whose third point, $XYZ$ (red), is skew-symmetric and thus lies off the quadric.
Let us consider, in analogy with the four-qubit case, the graph consisting of the three black vertices that are adjacent  to
$XXI$ (encircled gray), the three white vertices adjacent to $IZZ$ (also encircled gray) and the corresponding three inherited edges (gray parallel segments) -- which is the graph isomorphic to the Haar graph $H(4)$ (or, equivalently, to the point-point incidence graph of PG$(1,2)$). Joining the `red' point $XYZ$ with each of the three mid-points lying on  the edges of the $H(4)$ we get three out of 42 off-quadric lines of the hexagon (illustrated without third points by red segments). As there are $7 \times (7-3)= 28$
black-white vertex pairs of the above-defined type and  no two such pairs define the same off-quadric point, repeating this procedure we get all 28 off-quadric points of the hexagon and $28 \times 3/2 = 42$ remaining lines of the hexagon. It is also obvious that three lines issued from each `red' point lie in the same plane of  $\mathcal{W}(5,2)$, which means (see, e.\,g., \cite{Sanigaetal23}) that
a copy of the split Cayley hexagon we get by this construction is indeed {\it classically} embedded into $\mathcal{W}(5,2)$. 

\begin{figure}[t]
\centerline{\includegraphics[width=6.5truecm,clip=]{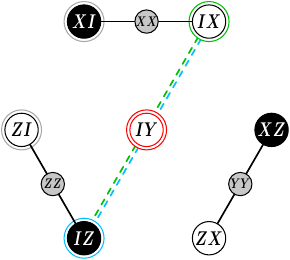}}
\caption{An illustration demonstrating that given a point-hyperplane incidence
graph of PG$(1,2)$ embedded into the two-qubit symplectic polar space, our procedure does not result in any other `red' line of the space. (Compare with Figure~\ref{3q-to-uns-full}.)}
\label{2q-to-uns-full}
\end{figure}

In the two-qubit case, our starting point is a
configuration isomorphic to the above-introduced $H(4)$-graph-underpinned configuration. This configuration comprises three pairwise
disjoint lines of the doily and lies on one of its hyperbolic quadrics, which is again taken to be the symmetric one, $\mathcal{Q}^{+}_{II}(3,2)$ -- as sketched in Figure~\ref{2q-to-uns-full}. As before, let us consider two non-adjacent vertices of the underlying $H(4)$ graph, one white
(e.\,g. $IX$ (green)) and one black (e.\,g. $IZ$ (blue)). They define an off-doily line
(double dashed) whose third point ($IY$ (red)) lies off the quadric. It is obvious that here we have only two different points that are adjacent to our selected points; namely, $XI$ (neighbor to $IX$) and $ZI$ (adjacent to $IZ$). Since these two vertices are {\it not} adjacent our procedure, in contrast with the above discussed two cases, ends here, thus not yielding any further line of the doily passing through the `red' point $IY$! And since the absence of `red' lines characterizes any off-quadric point in the doily, this finding can be rephrased by saying that a set of three lines of the doily associated with the three edges of an  $H(4)$ graph will always be the only unsatisfied contexts for the contextual configuration comprising all the 15 contexts of the two-qubit doily. This also provides a sort of explanation for the fact that ${F}^{{\tt uns}}_{2}$ does not cover all the points of the doily.
 
To conclude this subsection we return back to the central theme by stressing the following striking observation: whereas ${F}^{{\tt uns}}_{4} \cap \mathcal{Q}^{-}(7,2) \cong {E}^{{\tt uns}}_{4}$ for any $\mathcal{Q}^{-}(7,2) \in \mathcal{W}(7,2)$ (that is to say, criterion (\ref{cr-ell}) is satisfied),
${F}^{{\tt uns}}_{4} \cap \mathcal{Q}^{+}(7,2) \cong {H}^{{\tt uns}}_{4}$ holds just for a single $\mathcal{Q}^{+}(7,2) \in \mathcal{W}(7,2)$ (and so criterion (\ref{cr-hyp}) does not hold)!

\subsection{Contextuality in the five-qubit space}

The complexity of unsatisfied configurations we found here is much greater than in the previous case and so we will only describe their basic features.

\begin{figure}[pth!]
\centerline{\includegraphics[width=16.0truecm,clip=]{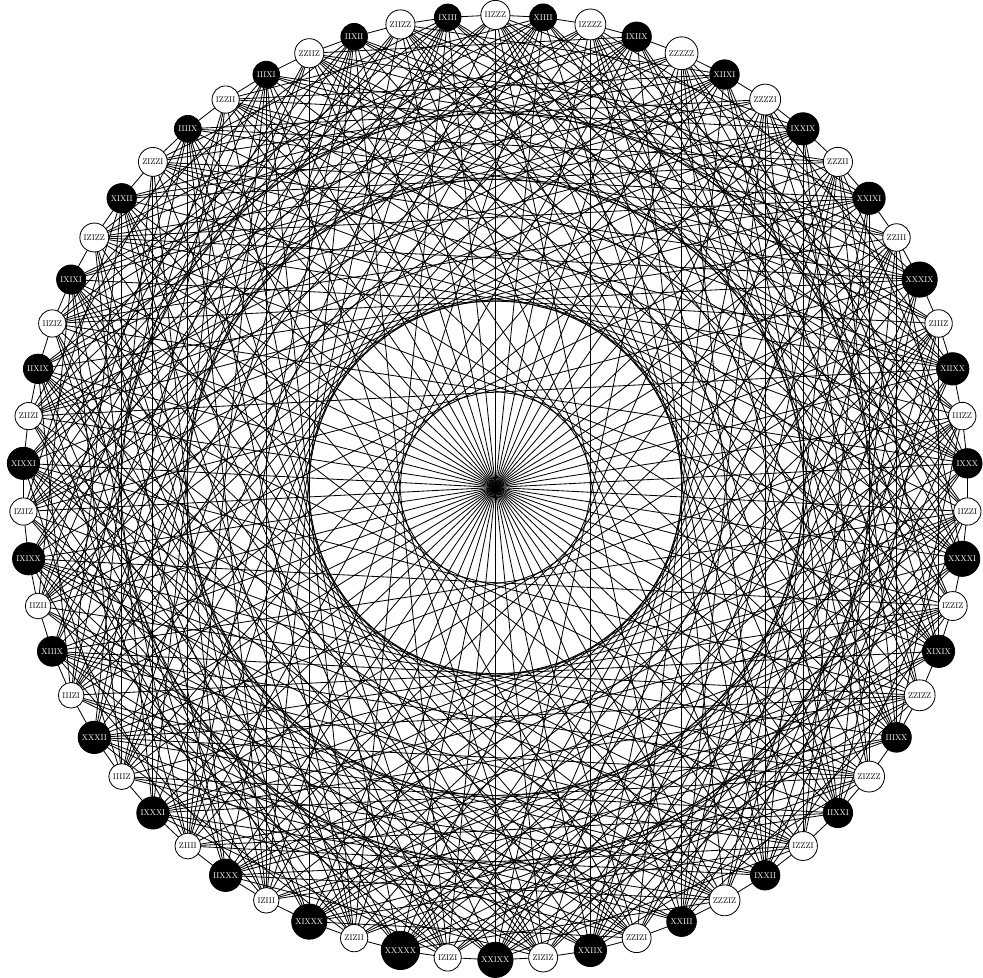}}
\vspace*{.2cm}
\caption{A graphical representation of the subgeometry formed by the 465 lines of the first type; in order not to make the figure to look much too crowded, the 465 points of the dash type are not shown.  The underlying point-hyperplane incidence graph of PG$(4,2)$ is rendered in the form isomorphic to the Haar graph $H(1103671145)$ whose number was computed for us by Dr. Eric W. Weisstein (Wolfram Research).}
\label{phig-pg42}
\end{figure}

We will start with hyperbolic quadrics. Such a quadric, $\mathcal{Q}^{+}(9,2)$, features 527 points and \np{23715} lines of which no less than \np{9420} can be negative. A related unsatisfied configuration we found contains only \np{6975} lines, so $d^{\rm hyp}_5 \leq 6975$ (see Table~\ref{resultsTable1} in Appendix~\ref{quadricDegreeApp}). This configuration exhibits a very high degree of symmetry as it has only two kinds of points and, similarly, two types of lines. Out of the 527 points, 62 are of degree 15 (solids) and 465 of degree 43 (dashes).
Out of the \np{6975} lines, there are 465 of type solid-solid-dash and the remaining \np{6510} ones are 
of type dash-dash-dash. Remarkably, one can associate the 62 solid points with the 62 vertices of the point-hyperplane incidence graph of PG$(4,2)$ in such a way that the 465
lines of the former type will be represented by the edges of this graph -- as illustrated
in Figure~\ref{phig-pg42}. If we compare this result with what we found for the four-qubit hyperbolic quadrics (see Figure~\ref{4q-ppig-pg32}) and for three-qubit ones (see Figure 2 in~\cite{muller2023new}), we arrive at the following natural conjecture: the core geometry of an unsatisfied configuration of a hyperbolic quadric $\mathcal{Q}^{+}(2N-1,2)$, $N \geq 3$, is underlined by the point-hyperplane incidence graph of the projective space PG$(N-1,2)$.

An elliptic quadric, $\mathcal{Q}^{-}(9,2)$,  is endowed with 495 points and \np{19635} lines, of which no less than 
\np{7860} can be negative. An upper bound we found in this case, for $d^{\rm ell}_5$, amounts to \np{7087}
(see Table~\ref{resultsTable1} in Appendix~\ref{quadricDegreeApp}). The corresponding configuration is already too complex to be described in sufficient detail. We only mention that
its points are of as many as 11 different degrees, all odd, the smallest being seven and the largest 51, with only two points being of the smallest degree.

In the case of the contextual configuration comprising all \np{86955} lines of the five-qubit space $\mathcal{W}(9,2)$, we found an unsatisfied configuration having 
\np{31479} lines ($d^{\rm full}_5 \leq 31479$, see Table~\ref{k1Table}), which is less than \np{35400}, the number of negative lines in this space. Among its points,
we find again 11 distinct degrees, the smallest being 15 and the largest one amounting to 105. Strikingly, among cardinalities of different degrees we spot some distinguished numbers that
occur in the three-qubit space. In particular, there are two sets of points of different degree having 288 elements either, which could be related to 288 Conwell heptads of $\mathcal{W}(5,2)$~\cite{sbhg21}. Further, there are 105 points of degree 99, this pointing out to the 105 lines located on a $\mathcal{Q}^{+}(5,2)$. These and several other intriguing observations will be treated in a separate paper.

\subsection{Contextuality in the six-qubit space}

To round off our exposition of illustrative examples, we will also briefly address the six-qubit space.

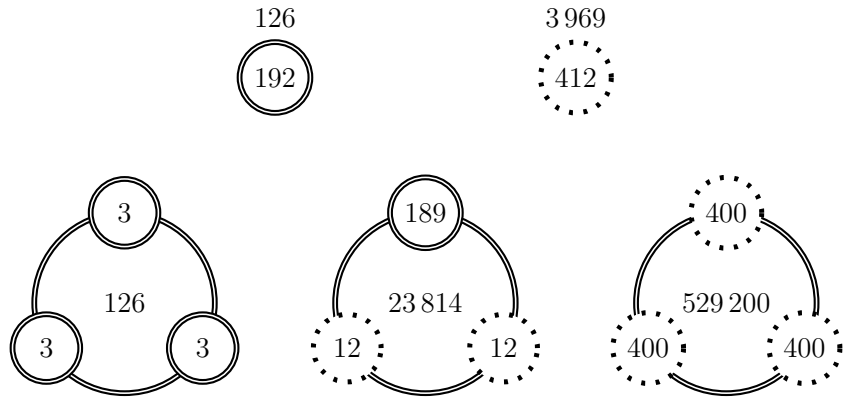
\begin{figure}[h]
\begin{center}
\begin{tikzpicture}[every plot/.style={smooth, tension=2},
  scale=1, 
  every node/.style={scale=0.9,fill=white,draw,circle,line width = 0.7mm,minimum size=1cm}
]
    \def\radius{1.2cm}
    \node[draw=none] at (2,3.8) {126}; 
    \node[double,thick] at (2,3) {192};
    
    \draw[double,thick] (0,0) circle (\radius);  
    \node[draw=none] at (0,0) {126}; 
    \node[double,thick] at (90:\radius) {3};
    \node[double,thick] at (210:\radius) {3};
    \node[double,thick] at (330:\radius) {3};
    
    \begin{scope}[shift={(4cm,0)}] 
        \node[draw=none] at (2,3.8) {\np{3969}}; 
        \node[dottednode] at (2,3) {412};
    
        \draw[double,thick] (0,0) circle (\radius);  
        \node[draw=none] at (0,0) {\np{23814}}; 
        \node[double,thick] at (90:\radius) {189};
        \node[dottednode] at (210:\radius) {12};
        \node[dottednode] at (330:\radius) {12};
    \end{scope}
    \begin{scope}[shift={(8cm,0)}] 
    
        \draw[double,thick] (0,0) circle (\radius);  
        \node[draw=none] at (0,0) {\np{529200}}; 
        \node[dottednode] at (90:\radius) {400};
        \node[dottednode] at (210:\radius) {400};
        \node[dottednode] at (330:\radius) {400};
    \end{scope}
\end{tikzpicture}
\end{center}

\caption{Properties of the point-line geometry comprising \np{553140} unsatisfied
constraints for the contextual geometry whose contexts are all the lines
of the space $\mathcal{W}(11,2)$, following the notation set up in Figure~\ref{4q-pattern-ell}.}
\label{6q-pattern-full}
\end{figure}

Here, our unsatisfied geometry associated with the contextual configuration comprising all \np{1396395} lines of  $\mathcal{W}(11,2)$ exhibits a great degree of combinatorial simplicity as it has only two different types of points and three distinct types of lines. In particular, out of the \np{4095} points of $\mathcal{W}(11,2)$, there are 126 of degree 192 (solids) and \np{3969} of degree 412 (dots), as portrayed in Figure~\ref{6q-pattern-full}, top layer. The totality of \np{553140} ($d^{\rm full}_6 \leq 553140$, much smaller than \np{615888}, the total number of negative contexts) unsatisfied lines features 126 lines
of type solid-solid-solid, \np{23814} lines of type solid-dot-dot and the remaining
\np{529200} ones consisting solely of dots -- see Figure~\ref{6q-pattern-full}, bottom layer. The 126 lines of the first type are quite remarkable as they split into two equally-sized disjoint sets, either of the two sets being isomorphic to nothing but a copy of the split Cayley hexagon of order two classically embedded into the subspace $\mathcal{W}(5,2)$ it spans, i.\,e. to an ${F}^{{\tt uns}}_{3}$.

In the unsatisfied configuration associated with a hyperbolic quadric of the space
we find as many as five different types of points and eight types of lines. Among its \np{2079} points, there are 14 of the smallest degree (31) and 49 of the largest one (211); moreover, the number of points having one specific degree (155) is equal to the number of linear doilies in  
$\mathcal{W}(5,2)$ -- 336~\cite{sbhg21}. Among its \np{132391} lines ($d^{\rm hyp}_6 \leq 132391$, see Table~\ref{resultsTable1} in Appendix~\ref{quadricDegreeApp}), there is a particular type of size 49  whose elements
can be associated with the edges of the complete bipartite graph $K_{7,7}$ ({\it aka} the adjacency graph of the Fano plane)\footnote{We note in passing that the Heawood graph is a subgraph of the $K_{7,7}$-graph.}; each such line consists of two points of the smallest degree and one point of the largest one.

Finally, the unsatisfied configuration we found for an elliptic quadric, comprising \np{131700} lines ($d^{\rm ell}_6 \leq 131700$, see Table~\ref{resultsTable1} in Appendix~\ref{quadricDegreeApp}), is the most complex of the three. The \np{2015} points of the quadric split into six different types and it is instructive to list them explicitly: 14 points of degree 84, 27 points of degree 92, 21 points of degree 100, 378 points of degree 196, \np{1008} points of degree 198 and 567 points of degree 204. The 27 points of degree 92 are found to be located on nine pairwise disjoint lines and so form a configuration isomorphic to ${E}^{{\tt uns}}_{3}$.
Moreover, the 14 points of degree 84 together with 21 points of degree 100 are situated on 21 lines that can be associated with the edges of the Heawood graph and so they form the configuration isomorphic to ${H}^{{\tt uns}}_{3}$. Finally, it is worth noticing that the number of points of degree 198 is the same as that of quadratic doilies living in $\mathcal{W}(5,2)$~\cite{sbhg21}, or as the number of ordinary hexagons contained in the split Cayley hexagon of order two.

\section{Conclusion}\label{sec:conc}

Making use of a new heuristic method in combination with symmetries exhibited by (specific subgeometries of) the multi-qubit symplectic polar space of order two $\mathcal{W}(2N-1,2)$, we were able, for $4 \leq N \leq 7$ qubits, to push the upper bounds on the degree of contextuality of contextual configurations living in
$\mathcal{W}(2N-1,2)$ much lower than those found with the previous method based on a SAT solver, and than the smallest number of negative lines a configuration of a given type can be endowed with. The power of this method is best illustrated with the four-qubit case, where we also achieved a deep, and fairly detailed, insight into the nature of those parts of contextual configurations that are irreproducible by any
NCHV model. The corresponding unsatisfied part of an elliptic quadric has in its core three copies of the split Cayley hexagon of order two sharing the Heawood graph and covering all the points of the quadric; this is quite a remarkable finding in light of the fact that the split Cayley hexagon of order two is the geometry that rules contextuality in the three-qubit $\mathcal{W}(5,2)$~\cite{Sanigaetal23}.  The unsatisfied contexts of a hyperbolic quadric are arranged into a structure isomorphic to $\mathcal{DW}(5,2)$, the space that is dual to $\mathcal{W}(5,2)$; here, we surmise to have reached also the lower bound. Interestingly, an unsatisfied copy of $\mathcal{DW}(5,2)$ also occurs in the case when all the lines of $\mathcal{W}(7,2)$ are considered as a contextual configuration, this time centered on a distinguished point-plane incidence graph of PG$(3,2)$ and surrounded by additional \np{1260} unsatisfied contexts. Hence, in all the three cases we see a clear connection with the {\it three}-qubit symplectic polar space $\mathcal{W}(5,2)$. Another remarkable fact in the four-qubit space is that two identical upper bounds (315) correspond to contextual configurations that are geometrically very different. Moreover, the upper bound we found for the contextuality degree of the whole space (\np{1575}) coincides with the number of lines on a hyperbolic quadric. Also, in the case of hyperbolic quadrics of three-, four- and five-qubit spaces we see an intriguing pattern where the core parts of the corresponding unsatisfied configurations are underpinned by the point-hyperplane incidence graphs of PG$(d,2)$, with $d = 2, 3$ and $4$, respectively.

Apart from dissecting in a similar fashion some higher-rank cases, we also plan to employ our new method to deal with contextual configurations whose contexts have more than three observables. For example, a natural context with four elements is isomorphic to an affine plane of order two, AG$(2,2)$. In $\mathcal{W}(5,2)$, there are already as many as 945 of them and it would be desirable to see what kind of contextual configurations they form, what the corresponding unsatisfied parts of them look like and how these are related to three-element-context configurations in this space. A well-known example of such a configuration is the so-called Mermin pentagram~\cite{mermin}. There are \np{12096} distinct pentagrams in $\mathcal{W}(5,2)$ and their properties have already been thoroughly analyzed~\cites{psh,ls,shj}. We would like to perform a similar study for other AG$(2,2)$-based classes of contextual configurations living in $\mathcal{W}(5,2)$. The largest of them is the $(63_{60}, 945_4)$-configuration that consists of all 945 AG$(2,2)$'s and its degree of contextuality should be equal to 189. We can arrive at this number in two different ways. One of them employs properties of the above-mentioned Mermin pentagrams. Thus, as each AG$(2,2)$ is contained in $12096 \times 5 / 945 = 64$ such pentagrams and the degree of contextuality of each pentagram is equal to one, we indeed get $12096 \times 1 / 64 = 189$. The other one makes use of the configuration comprising all 315 lines of
$\mathcal{W}(5,2)$, whose degree of contextuality amounts to 63~\cite{muller2023new}. Now, there are three planes through a line in $\mathcal{W}(5,2)$ and, so, there are three distinguished AG$(2,2)$s that we get by removing this common line from each of the three planes. Hence, it is natural to assume that to each of the 63 unsatisfied line contexts there will be three unsatisfied affine ones, which again yields $63 \times 3 = 189$. 

Another interesting task would be to explore those contextual configurations that
are common to two (or more) different $\mathcal{W}(2N-1,2)$'s living in the same ambient
space PG$(2N-1,2)$. The total number of (non-degenerate) symplectic polarities of rank $N$ in PG$(2N-1,2)$, $S_N$, is
given by the following formula (see, e.\,g., \cite{demb}, page 46)
\begin{equation*}
S_N = 2^{((2N-1)^2 -1)/4} \prod_{i=1}^{N-1} (2^{2i+1}-1).
\end{equation*}
Thus, we find that there are $S_2 = 2^2  (2^3 -1) = 28$ distinct doilies in PG$(3,2)$. The set of lines shared by two
different doilies is usually referred to as a \emph{linear congruence}. Such a congruence is rather simple, comprising either  a set of five pairwise disjoint lines forming a spread of PG$(3,2)$ (called an \emph{elliptic congruence}), or a set of six lines  lying in pairs in three planes meeting a distinguished line, the latter inclusive (a \emph{parabolic congruence};\footnote{For the sake of completeness, it is worth mentioning that in PG$(3,q)$, $q > 2$, there also exists a hyperbolic congruence, i.\,e. the congruence consisting of
$(q+1)^2$ lines incident to two skew lines, which is shared by $q-1$ different polarities.} see, e.\,g., \cite[Section 15.2]{hirsch-3d} or~\cite[$\S\S$ 8--17]{edge}.
However, in PG$(5,2)$ we already find as many as $S_3 = 2^6 (2^3 -1)(2^5-1) = 64 \times 7 \times 31 = 13888$ distinct $\mathcal{W}(5,2)$'s and so we expect  a greater variety and complexity of intersection patterns, which can still be tractable by our new method. 
On the other hand, we can -- for some small values of $N > 2$ -- consider all elliptic and hyperbolic quadrics contained in PG$(2N-1,2)$ and check quantum contextuality of their individual intersections with some selected $\mathcal{W}(2N-1,2)$ of the space.

\section*{Acknowledgments}
This work was supported in part by the Slovak VEGA grant agency, project number 2/0043/24,
by the EIPHI Graduate School (contract ``ANR-17-EURE-0002'') and by the Bourgogne-Franche-Comt\'e Region.
We would also like to cordially thank Dr. Petr Pracna (Technology Centre, Prague) for the help with Figures~\ref{hexagon}, \ref{4q-ppig-pg32}, \ref{DW52-210}, \ref{21pts} and~\ref{4q-to-uns-full}, Dr. Zsolt Szab\'o (Bosch Company, Budapest) for the help with Figures~\ref{3q-to-uns-full} and~\ref{2q-to-uns-full}, Dr. Henri de Boutray (ColibriTD, Paris) for creating for us the drawing of the Haar graph  $H(1103671145)$ as depicted in Figure~\ref{phig-pg42} and, especially, Dr. P\'eter Vrana (Budapest University of Technology and Economics, Budapest) for providing us with a particular labeling of the vertices of this graph in terms of five-qubit observables.

\bibliography{biblio}

\newpage
\appendix

\section{Exact values and upper bounds for the contextuality degree of quadrics}
\label{quadricDegreeApp}

\begin{table}[!htbp]
\begin{footnotesize}
\begin{center}
\begin{adjustbox}{scale=0.8}
\begin{tabular}{|l|c|r|r|r|r|l|l|}
\hline
Quadric Type       & $N$    & $p$ & $l$ &  $l^{-}$ & $|{K}^{-}|$            & $d$ & Duration\\
\hline
\hline
Hyperbolic          &  2     & 9      & 6          & 1 or 3         & 9 or 1                & 1 & 0\\
\hline
Hyperbolic          &  3     & 35     & 105        & 27 or 39       & 27 or 9               & 21 & 0\\
\hline
Hyperbolic          &  4     & 135    & \np{1575}  & 532 or 604 or 612& 81 or 54 or 1   & $\mathbf{\leq 315}$ & $<2$ s\\
\hline
Hyperbolic          & $\mathbf{5}$ & 527    & \np{23715} &\np{9420} or \np{9852} or \np{9900}& 243 or 270 or 15    & $\mathbf{\leq \np{6975}}$ & $<5$ s\\
\hline
Hyperbolic          & $\mathbf{6}$ & \np{2079}& \np{365211}    &\np{159376} or \np{161968} or \np{162256} or \np{162288}&  729 or \np{1215} or 135 or 1 & $\mathbf{\leq \np{132391}}$ & $<20$ s\\
\hline
Hyperbolic          & $\mathbf{7}$ & \np{8255}& \np{5720715}     & \np{2636592} or \np{2652144} or \np{2653872} or \np{2654064} &\np{2187} or \np{5103} or 945 or 21 & $\mathbf{\leq \np{2331191}}$ & $<10$ mn\\
\hline
Elliptic            &  2     & 5      & 0          & 0              & 6                     & N/A & 0\\
\hline
Elliptic            &  3     & 27     & 45         & 9 or 13        & 1 or 27                   & 9   & 0\\
\hline
Elliptic            &  4     & 119    & \np{1071}  & 360 or 384     & 12 or 108                  & $\mathbf{\leq 315}$  & $<2$ s\\
\hline
Elliptic            & $\mathbf{5}$ & 495    & \np{19635} &\np{7860} or \np{7876} or \np{8020} & 1 or 90 or 405 & $\mathbf{\leq \np{7087}}$ & $<5$ s\\
\hline
Elliptic            & $\mathbf{6}$ & \np{2015}& \np{332475}& \np{145920} or \np{146016} or \np{146880}&  18 or 540 or \np{1458} & $\mathbf{\leq \np{131700}}$ & $<20$ s\\
\hline
Elliptic           & $\mathbf{7}$ & \np{8127}& \np{5458635}& \np{2523024} or \np{2523088} or \np{2523664} or \np{2528848}& 1 or 189 or \np{2835} or \np{5103} & $\mathbf{\leq \np{2294580}}$ & $<10$ mn\\
\hline
\end{tabular}
\end{adjustbox}
\end{center}
\end{footnotesize}
\caption{Known results for the exact values and \textbf{new} results for upper bounds of the contextuality degree $d$ of quadrics. The notations are the same as in Table~\ref{k1Table} apart from the new symbol $|{K}^{-}|$ that stands for the number of quadrics having, respectively, a particular number of negative contexts listed in column $l^{-}$.
\label{resultsTable1}}
\end{table}
\newpage
\section{A particular bijection between three-qubit Fano planes and four-qubit observables}  \label{app}

An explicit form of the bijection between 135 planes (each listed as a set of points/obser\-vables) of the three-qubit $\mathcal{W}(5,2)$ and 135 points/observables of the four-qubit hyperbolic quadric $\mathcal{Q}_{IIII}^{+}(7,2)$ furnished by the $LGr(3,6)$, which was employed in Section~\ref{sec:res} to figure out basic properties of unsatisfied configurations of both a hyperbolic quadric
and the whole four-qubit symplectic polar space $\mathcal{W}(7,2)$. For the reader's convenience,  planes 1 to 105 are consecutively arranged into seven sets of 15 elements each, the planes in each such set passing through the same point/observable (listed first and  separated by a semicolon from the rest); the corresponding seven `first' observables pairwise anticommute and represent a particular Conwell heptad of $\mathcal{Q}_{III}^{+}(5,2)$ (for the definition of a Conwell heptad in the three-qubit setting, see, e.\,g., \cite{sbhg21}).
\begin{center}
\begin{tabular}{|r|l|c|}
\hline \hline
No. &  Fano plane in $\mathcal{W}(5,2)$   & Point on $\mathcal{Q}^+(7,2)$\\
\hline \hline
1   & $\{IIY; XYI, ZXY, XYY, ZXI, YZY, YZI \}$  & $YZXY$ \\
2   & $\{IIY; XYI, YXI, XYY, YXY, ZZI, ZZY \}$  & $ZYYX$ \\
3   & $\{IIY; XYI, IYI, XYY, IYY, XII, XIY \}$  & $XXZZ$ \\
4   & $\{IIY; XXY, ZYI, XXI, ZYY, YZY, YZI \}$  & $XYYZ$ \\ 
5   & $\{IIY; XXY, YYY, XXI, YYI, ZZI, ZZY \}$  & $IYYI$ \\
6   & $\{IIY; XXY, IXY, XXI, IXI, XII, XIY \}$  & $XIIZ$ \\
7   & $\{IIY; YII, IZY, YIY, IZI, YZY, YZI \}$  & $ZXZX$ \\
8   & $\{IIY; YII, YYY, YIY, YYI, IYY, IYI \}$  & $YYYY$ \\
9   & $\{IIY; YII, IXY, YIY, IXI, YXY, YXI \}$  & $XZXZ$ \\
10  & $\{IIY; ZIY, IZY, ZII, IZI, ZZI, ZZY \}$  & $ZIIX$ \\  
11  & $\{IIY; ZIY, ZYI, ZII, ZYY, IYY, IYI \}$  & $ZZXX$ \\
12  & $\{IIY; ZIY, IXY, ZII, IXI, ZXI, ZXY \}$  & $IZXI$ \\
13  & $\{IIY; XZY, IZY, XZI, IZI, XII, XIY \}$  & $IXZI$ \\
14  & $\{IIY; XZY, ZYI, XZI, ZYY, YXY, YXI \}$  & $YXZY$ \\
15  & $\{IIY; XZY, YYY, XZI, YYI, ZXI, ZXY \}$  & $YIIY$ \\
\hline
16  & $\{ZYX; YIZ, ZXY, XYY, IZZ, XXX, YZI \}$  & $IXYY$ \\
17  & $\{ZYX; YIZ, YXI, XYY, XZX, IXZ, ZZY \}$  & $YYIZ$ \\
18  & $\{ZYX; YIZ, IYI, XYY, ZIX, YYZ, XIY \}$  & $YZYX$ \\ 
19  & $\{ZYX; YZX, ZYI, XXI, IIX, XXX, YZI \}$  & $XZZI$ \\
20  & $\{ZYX; YZX, YYY, XXI, XIZ, IXZ, ZZY \}$  & $XXXZ$ \\
21  & $\{ZYX; YZX, IXY, XXI, ZZZ, YYZ, XIY \}$  & $IYYZ$ \\
22  & $\{ZYX; XYZ, IZY, YIY, ZXZ, XXX, YZI \}$  & $XYXY$ \\
23  & $\{ZYX; XYZ, YYY, YIY, XIZ, ZIX, IYI \}$  & $YIYI$ \\
24  & $\{ZYX; XYZ, IXY, YIY, ZZZ, XZX, YXI \}$  & $ZYZY$ \\
25  & $\{ZYX; IYX, IZY, ZII, ZXZ, IXZ, ZZY \}$  & $ZZXI$ \\
26  & $\{ZYX; IYX, ZYI, ZII, IIX, ZIX, IYI \}$  & $IZIX$ \\
27  & $\{ZYX; IYX, IXY, ZII, ZZZ, IZZ, ZXY \}$  & $ZIXX$ \\
28  & $\{ZYX; YXX, IZY, XZI, ZXZ, YYZ, XIY \}$  & $YXIY$ \\
29  & $\{ZYX; YXX, ZYI, XZI, IIX, XZX, YXI \}$  & $XIZX$ \\
30  & $\{ZYX; YXX, YYY, XZI, XIZ, IZZ, ZXY \}$  & $ZXZZ$ \\
  \hline \hline
\end{tabular}
\end{center}

\begin{center}
\begin{tabular}{|r|l|c|}
\hline \hline
No. &  Fano plane in $\mathcal{W}(5,2)$   & Point on $\mathcal{Q}^+(7,2)$\\
\hline \hline
31  & $\{YIX; ZYZ, ZXY, XYY, XXZ, IZX, YZI \}$  & $YYZI$ \\  
32  & $\{YIX; ZYZ, YXI, XYY, IXX, XZZ, ZZY \}$  & $XIYY$ \\
33  & $\{YIX; ZYZ, IYI, XYY, YYX, ZIZ, XIY \}$  & $ZYXY$ \\ 
34  & $\{YIX; YZX, YII, IZI, IIX, IZX, YZI \}$  & $IIZX$ \\
35  & $\{YIX; YZX, ZIY, IZI, XIZ, XZZ, ZZY \}$  & $ZXZI$ \\
36  & $\{YIX; YZX, XZY, IZI, ZZZ, ZIZ, XIY \}$  & $ZXIX$ \\ 
37  & $\{YIX; XYZ, XXY, ZYY, ZXZ, IZX, YZI \}$  & $YYIX$ \\
38  & $\{YIX; XYZ, ZIY, ZYY, XIZ, YYX, IYI \}$  & $YXYZ$ \\
39  & $\{YIX; XYZ, XZY, ZYY, ZZZ, IXX, YXI \}$  & $IZYY$ \\
40  & $\{YIX; IYX, XXY, YYI, ZXZ, XZZ, ZZY \}$  & $YXXY$ \\
41  & $\{YIX; IYX, YII, YYI, IIX, YYX, IYI \}$  & $XZZX$ \\
42  & $\{YIX; IYX, XZY, YYI, ZZZ, XXZ, ZXY \}$  & $ZYYZ$ \\
43  & $\{YIX; YXX, XXY, IXI, ZXZ, ZIZ, XIY \}$  & $IZXZ$ \\
44  & $\{YIX; YXX, YII, IXI, IIX, IXX, YXI \}$  & $XZII$ \\
45  & $\{YIX; YXX, ZIY, IXI, XIZ, XXZ, ZXY \}$  & $XIXZ$ \\
\hline
46  & $\{YZZ; ZYZ, ZYI, XXI, XXZ, IIZ, YZI \}$  & $IXXZ$ \\
47  & $\{YZZ; ZYZ, YYY, XXI, IXX, XIX, ZZY \}$  & $XZZZ$ \\
48  & $\{YZZ; ZYZ, IXY, XXI, YYX, ZZX, XIY \}$  & $XYYI$ \\
49  & $\{YZZ; YIZ, YII, IZI, IZZ, IIZ, YZI \}$  & $ZXII$ \\
50  & $\{YZZ; YIZ, ZIY, IZI, XZX, XIX, ZZY \}$  & $IXZX$ \\
51  & $\{YZZ; YIZ, XZY, IZI, ZIX, ZZX, XIY \}$  & $ZIZX$ \\ 
52  & $\{YZZ; XYZ, XYI, ZXI, ZXZ, IIZ, YZI \}$  & $ZIXZ$ \\
53  & $\{YZZ; XYZ, ZIY, ZXI, XZX, YYX, IXY \}$  & $YZIY$ \\
54  & $\{YZZ; XYZ, XZY, ZXI, ZIX, IXX, YYY \}$  & $XZXX$ \\
55  & $\{YZZ; IYX, XYI, YXY, ZXZ, XIX, ZZY \}$  & $XYIY$ \\
56  & $\{YZZ; IYX, YII, YXY, IZZ, YYX, IXY \}$  & $ZXYY$ \\  
57  & $\{YZZ; IYX, XZY, YXY, ZIX, XXZ, ZYI \}$  & $YZYI$ \\
58  & $\{YZZ; YXX, XYI, IYY, ZXZ, ZZX, XIY \}$  & $YYXX$ \\
59  & $\{YZZ; YXX, YII, IYY, IZZ, IXX, YYY \}$  & $IIYY$ \\
60  & $\{YZZ; YXX, ZIY, IYY, XZX, XXZ, ZYI \}$  & $YYZZ$ \\
\hline
61  & $\{XYX; ZYZ, IZY, YIY, XXZ, ZXX, YZI \}$  & $YZYZ$ \\
62  & $\{XYX; ZYZ, YYY, YIY, ZIZ, XIX, IYI \}$  & $IYIY$ \\
63  & $\{XYX; ZYZ, IXY, YIY, XZZ, ZZX, YXI \}$  & $YXYX$ \\
64  & $\{XYX; YIZ, XXY, ZYY, IZZ, ZXX, YZI \}$  & $ZIYY$ \\
65  & $\{XYX; YIZ, ZIY, ZYY, YYZ, XIX, IYI \}$  & $XYZY$ \\
66  & $\{XYX; YIZ, XZY, ZYY, IXZ, ZZX, YXI \}$  & $YYXI$ \\
67  & $\{XYX; YZX, XYI, ZXI, IIX, ZXX, YZI \}$  & $XZIX$ \\
68  & $\{XYX; YZX, ZIY, ZXI, YYZ, XZZ, IXY \}$  & $YIXY$ \\ 
69  & $\{XYX; YZX, XZY, ZXI, IXZ, ZIZ, YYY \}$  & $ZZXZ$ \\
70  & $\{XYX; IYX, XYI, XII, IIX, XIX, IYI \}$  & $XIZI$ \\
71  & $\{XYX; IYX, XXY, XII, IZZ, XZZ, IXY \}$  & $IXZZ$ \\
72  & $\{XYX; IYX, XZY, XII, IXZ, XXZ, IZY \}$  & $XXIZ$ \\
73  & $\{XYX; YXX, XYI, ZZI, IIX, ZZX, YXI \}$  & $IZZX$ \\ 
74  & $\{XYX; YXX, XXY, ZZI, IZZ, ZIZ, YYY \}$  & $ZXXX$ \\
75  & $\{XYX; YXX, ZIY, ZZI, YYZ, XXZ, IZY \}$  & $ZYYI$ \\
  \hline \hline
\end{tabular}
\end{center}

\begin{center}
\begin{tabular}{|r|l|c|}
\hline \hline
No. &  Fano plane in $\mathcal{W}(5,2)$   & Point on $\mathcal{Q}^+(7,2)$\\
\hline \hline
76  & $\{IYZ; ZYZ, IZY, ZII, IXX, ZXX, ZZY \}$  & $IZXX$ \\ 
77  & $\{IYZ; ZYZ, ZYI, ZII, ZIZ, IIZ, IYI \}$  & $ZIXI$ \\
78  & $\{IYZ; ZYZ, IXY, ZII, IZX, ZZX, ZXY \}$  & $ZZIX$ \\
79  & $\{IYZ; YIZ, XXY, YYI, XZX, ZXX, ZZY \}$  & $YZZY$ \\
80  & $\{IYZ; YIZ, YII, YYI, YYZ, IIZ, IYI \}$  & $ZXXZ$ \\
81  & $\{IYZ; YIZ, XZY, YYI, XXX, ZZX, ZXY \}$  & $XYYX$ \\
82  & $\{IYZ; YZX, XYI, YXY, XIZ, ZXX, ZZY \}$  & $YIYZ$ \\ 
83  & $\{IYZ; YZX, YII, YXY, YYZ, IZX, IXY \}$  & $YYZX$ \\ 
84  & $\{IYZ; YZX, XZY, YXY, XXX, ZIZ, ZYI \}$  & $IYXY$ \\
85  & $\{IYZ; XYZ, XYI, XII, XIZ, IIZ, IYI \}$  & $IXIZ$ \\
86  & $\{IYZ; XYZ, XXY, XII, XZX, IZX, IXY \}$  & $XXZI$ \\
87  & $\{IYZ; XYZ, XZY, XII, XXX, IXX, IZY \}$  & $XIZZ$ \\
88  & $\{IYZ; YXX, XYI, YZY, XIZ, ZZX, ZXY \}$  & $YXYI$ \\
89  & $\{IYZ; YXX, XXY, YZY, XZX, ZIZ, ZYI \}$  & $ZYIY$ \\
90  & $\{IYZ; YXX, YII, YZY, YYZ, IXX, IZY \}$  & $XZYY$ \\  
\hline
91  & $\{YXZ; ZYZ, IZY, XZI, YYX, ZXX, XIY \}$  & $YIZY$ \\ 
92  & $\{YXZ; ZYZ, ZYI, XZI, XZZ, IIZ, YXI \}$  & $ZXIZ$ \\
93  & $\{YXZ; ZYZ, YYY, XZI, IZX, XIX, ZXY \}$  & $XXZX$ \\
94  & $\{YXZ; YIZ, XXY, IXI, ZIX, ZXX, XIY \}$  & $XZXI$ \\
95  & $\{YXZ; YIZ, YII, IXI, IXZ, IIZ, YXI \}$  & $IIXZ$ \\
96  & $\{YXZ; YIZ, ZIY, IXI, XXX, XIX, ZXY \}$  & $XZIZ$ \\
97  & $\{YXZ; YZX, XYI, IYY, ZZZ, ZXX, XIY \}$  & $ZZYY$ \\
98  & $\{YXZ; YZX, YII, IYY, IXZ, IZX, YYY \}$  & $YYII$ \\
99  & $\{YXZ; YZX, ZIY, IYY, XXX, XZZ, ZYI \}$  & $XXYY$ \\
100 & $\{YXZ; XYZ, XYI, ZZI, ZZZ, IIZ, YXI \}$  & $ZXXI$ \\
101 & $\{YXZ; XYZ, XXY, ZZI, ZIX, IZX, YYY \}$  & $ZZZX$ \\ 
102 & $\{YXZ; XYZ, ZIY, ZZI, XXX, YYX, IZY \}$  & $IYYX$ \\
103 & $\{YXZ; IYX, XYI, YZY, ZZZ, XIX, ZXY \}$  & $IYZY$ \\
104 & $\{YXZ; IYX, XXY, YZY, ZIX, XZZ, ZYI \}$  & $YIYX$ \\
105 & $\{YXZ; IYX, YII, YZY, IXZ, YYX, IZY \}$  & $YYXZ$ \\
\hline
106 & $\{XXI, IIX, IXX, XXX, XIX, IXI, XII \}$  & $XIII$ \\ 
107 & $\{ZII, ZZZ, IZI, IZZ, ZZI, ZIZ, IIZ \}$  & $ZIII$ \\
108 & $\{ZXZ, YIY, XZZ, XXX, YYI, ZZX, IYY \}$  & $XXXX$ \\
109 & $\{XXZ, XIZ, YXY, IXI, ZIX, ZXX, YIY \}$  & $XIXI$ \\
110 & $\{ZZX, YYX, IXZ, XXI, ZYY, YZY, XIZ \}$  & $XXXI$ \\
111 & $\{ZXX, ZXI, YYI, IIX, XZX, XZI, YYX \}$  & $XIIX$ \\
112 & $\{YZY, YYZ, ZIX, IXX, XZZ, XYY, ZXI \}$  & $XIXX$ \\ 
113 & $\{XZI, IZX, ZYY, XIX, YXY, ZXZ, YYZ \}$  & $XXIX$ \\ 
114 & $\{XYY, XZX, YYX, IXZ, ZIZ, ZXI, YZY \}$  & $ZZIZ$ \\
115 & $\{ZXZ, XZZ, ZXI, YYI, IIZ, YYZ, XZI \}$  & $ZIIZ$ \\
116 & $\{XXZ, YXY, YYZ, ZIX, ZZI, IZX, XYY \}$  & $ZZZI$ \\
117 & $\{ZZX, IXZ, IZX, ZYY, ZII, IYY, ZXZ \}$  & $ZZII$ \\
118 & $\{ZXX, YYI, IYY, XZX, ZZZ, YIY, XXZ \}$  & $ZZZZ$ \\
119 & $\{YZY, ZIX, YIY, XZZ, IZI, XIZ, ZZX \}$  & $ZIZI$ \\
120 & $\{ZZZ, IZZ, IXX, ZII, ZYY, IYY, ZXX \}$  & $IIXX$ \\
121 & $\{IZI, ZIZ, XIX, ZZZ, XZX, YIY, YZY \}$  & $IXIX$ \\
  \hline \hline
\end{tabular}
\end{center}

\begin{center}
\begin{tabular}{|r|l|c|}
\hline \hline
No. &  Fano plane in $\mathcal{W}(5,2)$   & Point on $\mathcal{Q}^+(7,2)$\\
\hline \hline
122 & $\{IZZ, IIZ, XII, IZI, XZZ, XIZ, XZI \}$  & $IXII$ \\
123 & $\{ZIZ, ZZI, XXX, IZZ, YXY, YYX, XYY \}$  & $IXXX$ \\
124 & $\{IIZ, ZII, IXI, ZIZ, IXZ, ZXI, ZXZ \}$  & $IIXI$ \\
125 & $\{ZZI, ZZZ, XXI, IIZ, YYI, YYZ, XXZ \}$  & $IXXI$ \\
126 & $\{IIX, XXI, YYX, XXX, YYI, ZZX, ZZI \}$  & $IZZI$ \\
127 & $\{IXX, IIX, ZXI, IXI, ZIX, ZXX, ZII \}$  & $IZII$ \\ 
128 & $\{XIX, IXX, YYZ, XXI, ZYY, YZY, ZZZ \}$  & $IZZZ$ \\
129 & $\{XII, XIX, IZX, IIX, XZX, XZI, IZI \}$  & $IIZI$ \\
130 & $\{XXX, XII, IYY, IXX, XZZ, XYY, IZZ \}$  & $IIZZ$ \\
131 & $\{IXI, XXX, YIY, XIX, YXY, ZXZ, ZIZ \}$  & $IZIZ$ \\
132 & $\{XYY, IYY, XZX, XII, IXZ, XXZ, IZX \}$  & $XXII$ \\
133 & $\{XZI, ZYY, XIZ, YXY, IZZ, YYX, ZXX \}$  & $ZIZZ$ \\
134 & $\{ZII, IZI, IIX, ZZI, ZIX, IZX, ZZX \}$  & $IIIX$ \\
135 & $\{XXI, IXI, XIZ, XII, IXZ, XXZ, IIZ \}$  & $IIIZ$ \\
 \hline \hline
\end{tabular}
\end{center}

\end{document}